\documentclass[twocolumn]{aastex62}
\pdfoutput=1 %for arXiv submission
\usepackage{amsmath,amstext}
\usepackage[T1]{fontenc}
\usepackage{apjfonts}
\usepackage[figure,figure*]{hypcap}
\usepackage[nameinlink]{cleveref}
\makeatletter
\usepackage{etoolbox}
\usepackage{appendix}
\patchcmd\H@refstepcounter{\protected@edef}{\protected@xdef}{}{}
\makeatother
\usepackage{amsmath,amssymb}
\newlength{\colwidth}
\definecolor{bleudefrance}{rgb}{0.19, 0.55, 0.91}

\newcommand{\addgals}{\textsc{Addgals}}
\newcommand{\nbody}{$N$-body}

\newcommand{\hMsun}{{h^{-1}}{\rm M}_{\solar}}
\newcommand{\hmsun}{{h^{-1}}{\rm M}_{\solar}}
\newcommand{\msol}{\ensuremath{{\rm M}_{\solar}}}
\newcommand{\solar}{_{\mathord\odot}}
\newcommand{\hmpc}{\ifmmode{h^{-1}{\rm Mpc}}\;\else${h^{-1}}${\rm Mpc}\fi}
\newcommand{\hMpc}{\ifmmode{h^{-1}{\rm Mpc}}\;\else${h^{-1}}${\rm Mpc}\fi}
\newcommand{\hGpc}{\ifmmode{h^{-1}{\rm Gpc}}\;\else${h^{-1}}${\rm Gpc}\fi}
\newcommand{\hkpc}{\ifmmode{h^{-1}{\rm kpc}}\;\else${h^{-1}}${\rm kpc}\fi}

\newcommand{\msun}{{\rm M}_{\solar}}

\newcommand{\beq}{\begin{equation}}
\newcommand{\eeq}{\end{equation}}

\newcommand{\mr}{\ifmmode{M_r}\;\else$M_r$\fi}
\newcommand{\mrh}{\ifmmode{M_r - 5{\rm log}h}\;\else$M_r - 5{\rm log} h$\fi}
\newcommand{\rd}{\ifmmode{R_\delta}\;\else$R_\delta$\fi}
\newcommand{\ngals}{\ifmmode{N_{\rm gals}}\;\else$N_{\rm gals}$\fi}

\newcommand{\be}{\begin{equation}}
\newcommand{\ee}{\end{equation}}
\newcommand{\bee}{\begin{eqnarray}}
\newcommand{\eee}{\end{eqnarray}}

\newcommand{\hinv}{h^{-1}}

\newcommand{\ltsima}{$\; \buildrel < \over \sim \;$}
\newcommand{\gtsima}{$\; \buildrel > \over \sim \;$}
\newcommand{\gsim}{\lower.5ex\hbox{\gtsima}}
\newcommand{\lsim}{\lower.5ex\hbox{\ltsima}}

\newcommand{\Rvir}{\ifmmode{R_{\rm vir}}\else$R_{\rm vir}$\fi}
\newcommand{\rvir}{\ifmmode{R_{\rm vir}}\else$R_{\rm vir}$\fi}
\newcommand{\Mvir}{\ifmmode{M_{\rm vir}}\else$M_{\rm vir}$\fi}
\newcommand{\mvir}{\ifmmode{M_{\rm vir}}\else$M_{\rm vir}$\fi}

\shortauthors{Wechsler, DeRose et al.}
\shorttitle{\textsc{Addgals} Synthetic Sky Surveys}

\begin{document}

\title{ADDGALS: Simulated Sky Catalogs for Wide Field Galaxy Surveys}

\author[0000-0003-2229-011X]{Risa H. Wechsler}
\affiliation{Department of Physics, Stanford University, 382 Via Pueblo Mall, Stanford, CA 94305, USA}
\affiliation{Kavli Institute for Particle Astrophysics \& Cosmology, P. O. Box 2450, Stanford University, Stanford, CA 94305, USA}
\affiliation{SLAC National Accelerator Laboratory, Menlo Park, CA 94025, USA}
\author[0000-0002-0728-0960]{Joseph DeRose}
\affiliation{Lawrence Berkeley National Laboratory, 1 Cyclotron Road, Berkeley, CA 93720, USA}
\author{Michael T. Busha}
\affiliation{Kavli Institute for Particle Astrophysics \& Cosmology, P. O. Box 2450, Stanford University, Stanford, CA 94305, USA}
\affiliation{Present Address: Securiti}
\author[0000-0001-7774-2246]{Matthew R. Becker}
\affiliation{High-Energy Physics Division, Argonne National Laboratory, Lemont, IL 60439, USA}
\author{Eli Rykoff}
\affiliation{Kavli Institute for Particle Astrophysics \& Cosmology, P. O. Box 2450, Stanford University, Stanford, CA 94305, USA}
\affiliation{SLAC National Accelerator Laboratory, Menlo Park, CA 94025, USA}
\author{August Evrard}
\affiliation{Departments of Physics and Astronomy, University of Michigan, Ann Arbor, MI}

\begin{abstract}
We present a method for creating simulated galaxy catalogs with realistic galaxy luminosities, broad-band colors, and projected clustering over large cosmic volumes. The technique, denoted \addgals\ (Adding Density Dependent GAlaxies to Lightcone Simulations), uses an empirical approach to place galaxies within lightcone outputs of cosmological simulations. It can be applied to significantly lower-resolution simulations than those required for commonly used methods such as halo occupation distributions, subhalo abundance matching, and
semi-analytic models, while still accurately reproducing projected galaxy clustering statistics down to scales of $r\sim 100\, \hkpc$. We show that \addgals\ catalogs reproduce several statistical properties of the galaxy distribution as measured by the Sloan Digital Sky Survey (SDSS) main galaxy sample, including galaxy number densities, observed magnitude and color distributions, as well as luminosity- and color-dependent clustering. We also compare to cluster--galaxy cross correlations, where we find significant discrepancies with measurements from SDSS that are likely linked to artificial subhalo disruption in the simulations. Applications of this model to simulations of deep wide-area
photometric surveys, including modeling weak-lensing statistics, photometric redshifts, and galaxy cluster finding are presented in \citet{DeRose2018}, and an application to a full cosmology analysis of Dark Energy Survey (DES) Year 3 like data is presented in \citet{DeRose2021}.  We plan to publicly release a 10,313 square degree catalog constructed using \addgals\ with magnitudes appropriate for several existing and planned surveys, including SDSS, DES, VISTA, WISE, and LSST.
\end{abstract}

\section{Introduction}
Cosmology and the study of galaxy formation are undergoing a renaissance driven
by exponential increases in computing power, the public availability of large
amounts of high-quality sky survey data, and continued investment in ever-more sensitive
instrumentation. These trends place stringent demands on the accuracy of 
the theoretical models used to analyze such survey data.

The best current models of the Universe posit that hierarchical structure formation via
the gravitational collapse of cold dark matter drives the formation and evolution
of galaxies. Simultaneously, galaxies serve as a rich set of tracers of the cosmic density 
and velocity fields, imparting the galaxy distribution with sensitivity to 
fundamental physics like cosmic acceleration, modifications to General Relativity, 
massive neutrinos, and the micro-physical nature of dark matter \citep[e.g.,][]{weinberg_etal:12}. This rich discovery potential demands precise 
connection between the galaxy and matter distributions, particularly on small cosmic scales which hold immense statistical information.  The galaxy--dark matter connection is a key source of theoretical uncertainty in galaxy survey analyses concerned with constraints on fundamental physics.

The context above is driving two important trends in cosmology. First, researchers are
developing a wealth of methods that aim to infer the connection between galaxies
and their dark matter halos (see \citealt{WechslerTinker} for a review). These studies
usually employ large-volume, high-resolution $N$-body simulations of structure formation.
Second, cosmologists now routinely employ "synthetic" or "mock" catalogs of galaxies to
support analyses of survey data. These synthetic catalogs are constructed with a wide
variety of techniques that draw from advances in understanding the connection between
galaxies and halos.  Published examples focusing on modeling galaxy populations
in realistic survey lightcones include approaches using halo occupation distributions (HOD) \citep{yan_etal:04,manera_etal:13, sousbie_etal:08, Fosalba2015a, Crocce2015, Smith2016, Harnois-Deraps2018, Stein2020}, 
semi-analytic models (SAM) \citep{eke_etal:04, cai_etal:09, merson_etal:13, Somerville2021}, 
subhalo abundance matching (SHAM)
\citep{gerke_etal:12, Safonova2020} 
or a combinations of the above \citep{korytov2019} to accomplish this task.

In this work, we present \addgals\ (Adding Density-Determined Galaxies to Lightcone Simulations),
a computationally inexpensive, but high-fidelity approach for constructing synthetic galaxy 
catalogs from lightcone simulations, designed to support the analysis of large-area galaxy survey data. 
Unlike HOD, SAM, and SHAM approaches, it is designed specifically to populate
modest resolution $N$-body simulations with galaxies that have realistic luminosities, 
spectral energy distributions (SEDs), and clustering. Except for a few percent of galaxies occupying the most massive halos, \addgals\ is not sensitive to these N-body simulations' lack of convergence at the smallest scales.  The modest expense 
of these simulations enables the creation of large numbers of large-volume realizations of the Universe,
which are often required by modern survey analyses.

With its modest computational requirements, \addgals\ can be used to bring a new level of
realism to survey analysis tasks that require a \textit{statistical sample} of synthetic
catalogs. Examples of these tasks include generating covariance matrices 
or testing the robustness of these analyses to key systematic effects through direct, end-to-end tests
where the true answer is known. \citet{MacCrann17},  and \citet{DeRose2021b} present key examples of the latter approach. These works combine the Dark Energy Survey (DES) year one (Y1) and year three (Y3)
analysis pipelines with 18 synthetic catalogs produced with the methodology presented
in this work to perform end-to-end tests of a $3\times2$-point weak lensing and galaxy 
clustering analysis; \citet{To2021} performed a similar analysis that combined these statistics with cluster counts and cluster--galaxy cross correlations.
Due to the realism of the \addgals\ catalogs, they were able to use 
the same analysis pipeline as was applied to the DES data. These tests depended critically on simulated catalogs that jointly modeled several effects, including various observables (e.g. galaxy
clustering, galaxy-galaxy lensing, cosmic shear, and cluster counts), photometric redshifts, and survey effects like
varying depth maps. Further, tens of realizations were needed to demonstrate that the
recovered parameters were accurate to well below the sensitivity of the DES measurements
themselves. These kinds of tests will be increasingly important as surveys begin to produce
stringent constraints on fundamental physics, motivating an approach that is able to model
large volumes and multiple observables with modest computational cost.

As emphasized by \cite{WechslerTinker}, currently used approaches for modeling galaxies within
large-scale structure face trade offs between the fidelity of the modeled properties, the
resolution or computational requirements of the method, and the degree to which the model
is physics-driven or empirically data-driven. For example, the subhalo abundance matching approach (SHAM),
which assumes that all galaxies are placed on resolved halos and subhalos, has been shown to
faithfully reproduce the spatial distribution of galaxies in the local Universe where it can
best be measured \citep{Kravtsov04,Reddick12,ChavesMontero2016,Lehmann2017,Contreras2020,DeRose2021}, as well
as the evolution of the galaxy population with time \citep{Conroy06, Moster11, Behroozi12},
with a very small number of parameters. This technique has the advantage of including
several important correlations between halo history, galaxy populations, and environment
that are neglected by some other methods, but has stringent resolution requirements.  

One of the most commonly used methods, populating simulations with galaxies using a halo
occupation distribution \citep[HOD; e.g.][]{Jing1998, Seljak2000, Bullock2003, Yang2003, Berlind2002, Zheng2005, Mandelbaum2006, vandenBosch2007, Zehavi2011, Zu2015}, places each galaxy 
in regions within resolved {\em host} halos, irrespective of dark matter substructures.
This reduces the computational requirements on the simulations compared with methods
that trace halo histories, but generally requires more parameters than abundance matching 
and may be missing relevant correlations between galaxy populations and halo history.  The 
conditional luminosity function method \citep[e.g.][]{Yang03, Cooray2006} has similar
requirements. The computational expense of HOD modeling can be further decreased by 
employing approximate, or low-resolution methods for generating halo catalogs
\citep[e.g.][]{Bond96b, Scoccimarro2002, Kitaura2016, Chuang2017, Monaco2013, Tassev2013, White2014, Avila2015, Feng2016, Izard2018, Balaguera-Antolinez2018}.

SAMs \citep{White1991, Kauffmann1993, Somerville1999, Cole2000, Benson02, Bower2006, Benson2012, Guo2013, Croton2016}
generally require a degree of resolution between abundance matching and the HOD --- the former
is more relevant if one wants to trace the histories of all galaxies properly and if
one wants to keep every galaxy on a resolved substructure.  This can be reduced to
the less demanding requirements of the HOD if semi-analytic methods are
used to track halo histories and the kinematics of satellite galaxies in larger systems \citep{Benson2012,Jiang2016,Yang2020,Jiang2021}.

Concurrent with the development and use of the method described here, 
significant progress has been made in other data-driven
approaches, particularly in empirical models that use information from
halo histories. These approaches include extensions to the abundance matching
approach like conditional abundance matching, which associates color 
or star formation rates with secondary halo properties 
\citep{Masaki2013, Hearin2013a, Hearin2013b, Yamamoto2015, Saito2016, Contreras2020} 
or that trace galaxy histories through the histories of their halos
\citep[e.g.][]{Becker2015, Moster2018, Behroozi2018} and constrain their 
properties with a wide range of data.

Finally, full-physics cosmological hydrodynamics methods \citep[see][for a recent review]{Vogelsberger2020} are making steady progress in describing the galaxy--halo connection.  A recent verification of three independent simulations examines the satellite galaxy occupation conditioned on total halo mass and redshift, finding a consistent form for the probability density function along with slightly super-Poisson dispersion, but with mean counts varying by tens of percent \citep{Anbajagane2020}. These simulations are computationally expensive and thus challenging to use to model large survey volumes, but they can be used to inform HOD approaches and test SAM or empirical methods, and provide essential input into possible modification of the dark matter distribution from baryonic processes.

\addgals' combination of realism and relatively low computational expense owes to a machine-learning style 
approach that uses higher-resolution \nbody\ simulations to train the galaxy--dark matter connection scales and data to train a physically motivated model for the dependence of galaxy properties on local density.
This approach is similar in spirit to other recent work that employs statistical
learning techniques to connect the dark matter distribution to the distribution of biased tracers 
\citep{Modi2018,Berger2019,Ramanah2019,Zhang2019,Troster2019,Dai2020}. The features that 
\addgals\ uses are chosen to be relatively insensitive to resolution effects in the simulations, while still encapsulating quantities relevant to the physics of galaxy formation. 

\begin{figure*}
\center{
\includegraphics[width=0.82\textwidth]{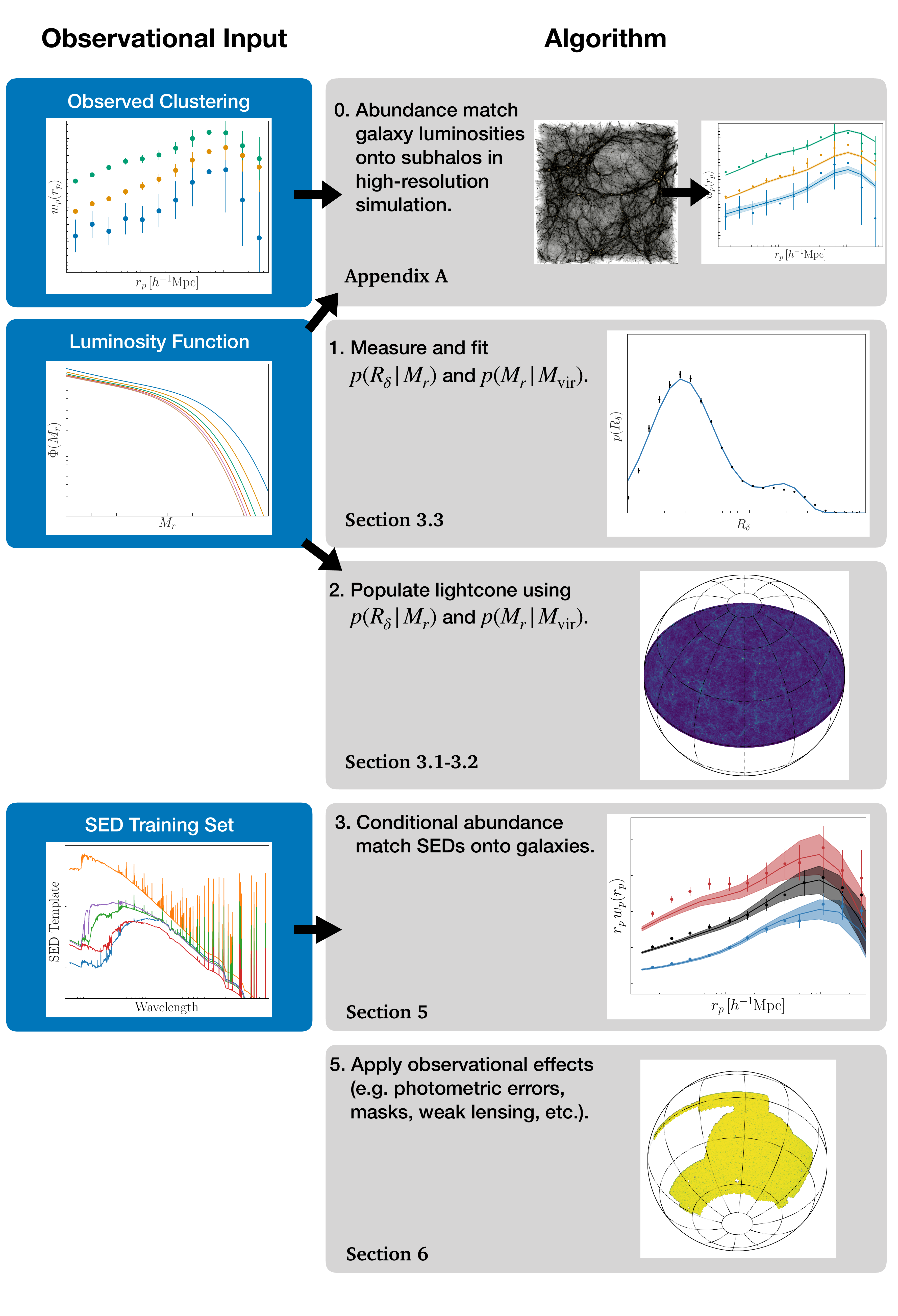}}
\caption{Flowchart of the \addgals\ algorithm.  Observational inputs are listed in the left hand column.  In the first step, we use observed clustering and luminosity functions to constrain a SHAM model, applied to a simulation with resolved substructures (\cref{app:density-sampling}).  In the second step, we measure and fit a model for central galaxies given halo mass (\cref{sec:cenpop}) and for the dark matter density \rd\ given luminosity for all other galaxies (\cref{sec:pop}).  In the third step, we populate a lightcone using this algorithm.  In the fourth step, we use an observed galaxy sample with luminosities, and SED properties to conditional abundance match SEDs onto simulated galaxies (\cref{sec:addcols}). Finally, we apply observational effects (\cref{sec:observe}).}
\label{fig:flowchart}
\end{figure*}

A flowchart with the key steps of the algorithm is given in \cref{fig:flowchart}.
The \addgals\ algorithm can be divided into two main parts, the assignment of luminosities and the assignment of SEDs. In the first part, we fit a model to the distribution of galaxy absolute magnitude at fixed local overdensity using a high-fidelity model of the galaxy--halo connection. In this work, we use a SHAM model applied to high-resolution structure formation simulations, but this choice is not essential. We then use these distributions to populate a low-resolution simulation via Monte Carlo. This process is illustrated in steps 1 through 3 in the flowchart. In the second part of the \addgals\ algorithm, we use a conditional abundance matching model fit to the Sloan Digital Sky Survey (SDSS) in \citet{DeRose2021} to assign an SED to each galaxy. Finally, we apply observational effects to produce the complete catalog. These parts correspond to steps 4 and 5 in the flowchart.

We demonstrate that these steps are able to reproduce the absolute magnitude dependent
two-point clustering and halo occupation properties of the SHAM catalog. Further, 
we show that they reproduce a number of additional observed properties of SDSS galaxies,
including their color distributions at a given absolute magnitude and the qualitative
trends of the observed color-dependent clustering.

The model that is perhaps most similar to the one presented in this work is \textsc{GalSampler}
\citep{Hearin2019}, in that it places galaxies from a high-fidelity model of galaxy 
formation run on high-resolution simulations into the halos of
lower resolution simulations. The main distinguishing factor is the use of halo mass as
the conditional variable in \textsc{GalSampler}, whereas \addgals\ uses local Lagrangian density,
which can be measured in significantly lower-resolution simulations.

In this work, \addgals\ is trained on a SHAM model, but the machine-learning style 
approach taken by \addgals\ generalizes to training on other models of the 
galaxy-halo connection, including hydrodynamical simulations, SAMs,
or empirical models that trace halo histories. Note that it is likely that secondary 
properties of the density field will be needed for these generalizations in analogy
to secondary halo properties and assembly bias. This flexibility combined with modest
computational requirements will enable the production of suites of synthetic catalogs with different 
underlying models for the galaxy--halo connection. These suites can then be used to test
the robustness of cosmological constraints from surveys to underlying assumptions about
galaxy formation.

\addgals\ has been in use for some time to facilitate a variety of applications of 
large-scale sky survey data, with a particular focus on wide-area photometric surveys. 
A preliminary description of the work was presented in \cite{wechsler:04}. Subsequent 
work using these catalogs has made use of earlier versions than those described here; 
in most cases the important details were described in those papers. Because of the
ability of these techniques to accurately model large, wide surveys including realistic
photometry and lensing, the range of applications has been broad.  

Catalogs produced with this method have been used extensively in the testing, systematics assessment,
and co-analysis of  galaxy cluster catalogs and results with the \textsc{maxBCG}
and \textsc{redMaPPer} algorithms
\citep{koester_etal:07, koester_etal:07b, johnston_etal:07a, rozo_etal:07, rozo_etal:07b, becker_etal:07, sheldon_etal:09b, hansen_etal:09, tinker_etal:11, dietrich_etal:14, Farahi2014, Varga2018, Abbott2020, To2021, Myles2021},
for the improvement and testing of other galaxy cluster finders
\citep{miller_etal:05, dong_etal:08, Hao10, soares-santos_etal:11, Bleem2015},
for the development and testing of a number of photometric redshift algorithms,
especially in the context of DES
\citep{Gerdes10, Cunha12, Cunha12b, Bonnett15, Leistedt15, Hoyle17, gatti17, Cawthon18, Buchs2019, Myles2020b, Gatti2020, Cawthon2020} and LSST \citep{Malz18, Schmidt20},
for the development of various survey analysis approaches using weak lensing shear
\citep{vanderplas2012, Chang2014, Szepietowski2014, Becker2016, Troxel2018, Freidrich2018, Chang2018, Bradshaw2019}, 
for the development and systematics testing of various galaxy and cluster cross correlations
\citep{High2012, Bleem2012, Shin2018, Pandey19} and other statistics of the galaxy and lensing spatial distribution
\citep{Friedrich18, Gruen18}, and for early preparation and testing of
the science prospects of the DES \citep{Gill12, Davies:12, Chang15, Park16, Asora16}, 
the WFIRST survey \citep{Martens2018, Massara2020}, and the LSST survey \citep{Mao2018} as 
well as spectroscopic surveys \citep{Saunders14, Nord16}.

In \cite{DeRose2018}, we described the use of \addgals\ to create
a suite of synthetic catalogs for the DES, extending the work described here to higher
redshift while including a number of additional observational effects and presenting tests of a number of additional observables, including those related to cosmic shear, photometric redshifts, high redshift galaxy clustering and lensing, 
and photometric cluster finding. That suite
of catalogs as well as earlier versions were used extensively in the analysis of
DES Science Verification and Y1 data 
\citep{y1kp, Krause17, MacCrann17, Gruen17, Sanchez17, Clampitt17, Davis17, Cawthon18, Varga18, Gatti18, Chang18, Abbott20}.
Catalogs produced using \addgals\ have continued to be used to facilitate DES Y3 cosmology analyses 
\citep{Buchs2019, Myles2020b, Gatti2020, Cawthon2020}, and a description of the catalogs used for that work is given in \cite{DeRose2021b}.
 
This paper proceeds as follows. In \cref{sec:sims}, we describe the simulations
used in this work. In \cref{sec:addlums} our method of populating simulations with galaxies in a single-band rest-frame absolute magnitude is described. Tests of this part of the algorithm are presented in \cref{sec:lumtests}. In \cref{sec:addcols}, we
outline our method for assigning spectral energy distributions to simulated galaxies.
Tests of this method are presented in \cref{sec:coltests}. In \cref{sec:res_requirements}
we discuss the resolution requirements for \addgals. Finally, we conclude in
\cref{sec:conclusions} with a discussion of the strengths and limitations of the algorithm and future directions of research. Throughout this manuscript, we quote magnitudes
using the AB system and $h = 1.0$ units.

\section{$N$-body simulations}\label{sec:sims}

All simulations in this work were run using the code \textsc{L-Gadget2} \citep{springel_etal:05}, a proprietary version of \textsc{Gadget-2} optimized for memory efficiency 
and explicitly designed to run large-volume, dark matter-only $N$-body simulations. 
We have modified this code to create a particle lightcone output on the fly \citep[see][for details]{DeRose2018}. Initial conditions were generated with the second-order Lagrangian perturbation theory code \textsc{2LPTic} \citep{crocce_etal:06}  using linear power spectra 
computed with the \textsc{CAMB} code \citep{lewis:04}. Early versions of these simulations were generated on XSEDE supercomputers using the Apache Airavata\footnote{https://airavata.apache.org/}  workflow management framework \citep{Erickson:2012:HTW:2335755.2335830}.

We use four $N$-body simulations with volumes of $(400\, \hMpc)^3$, $(1.05\, \hGpc)^3$, $(2.6\, \hGpc)^3$, and $(4.0\, \hGpc)^3$; the simulation parameters are summarized in \cref{table:simulations}. The first of these, deemed \textsc{T1} (Training Simulation 1), requires sufficient resolution that 
a SHAM approach can reasonably model the galaxy distribution down
to roughly $M_r = -19$ (see e.g.\citealt{Reddick12,Lehmann2017})  This simulation has $L_{\rm box}=400~\hinv$Mpc and $2048^3$ particles. At this resolution, the SHAM catalog is not strictly complete down to $M_r=-19$, as subhalos
that would host galaxies with $M_{r}<-19$ near the cores of massive hosts become stripped and are no longer trackable by the halo finder as they have too few particles (see \citealt{Reddick12} for a detailed discussion). However, comparisons with SDSS data show that the resolution 
is sufficient to model the observed two-point function within current observational constraints down to $M_r = -19$, except on the very smallest scales for dimmest galaxies in this sample. It also does reasonably well for galaxies fainter than this limit \citep[see][for further discussion]{DeRose2021}. The inability to accurately model small-scale clustering of the faintest samples owes in large part to subhalo disruption in \textsc{T1} simulation. This is discussed at greater length in \cref{sec:res_requirements}, where we compare with the \textsc{C250} simulation, run with the same settings as the \textsc{T1} simulation, but in a volume of $(250\, h^{-1}\rm Mpc)^3$, using $2560^3$ particles, and a force softening of $\epsilon=0.8\hkpc$.
A lightcone output is not necessary for the \textsc{T1} simulation, but merger trees are required to construct the abundance matching catalog. We save 100 simulation snapshots logarithmically spaced from $z = 12$ to $z = 0$, which allows for the construction of accurate merger trees. 

For the three larger simulations, \textsc{L1}, \textsc{L2}, and \textsc{L3} (Lightcone Simulations 1--3), ten snapshots at redshifts $$z=\{0.0, 0.10, 0.25, 0.4, 0.5, 0.7, 0.85, 1, 2, 3\}$$ are produced, as well as lightcones with areas of $10,313$ square degrees each (one quarter of the sky). The \textsc{L1} simulation is used to produce the simulated galaxy catalog that we compare to SDSS in \cref{sec:coltests}, while \textsc{L2} and \textsc{L3} are used solely for the resolution tests presented in \cref{sec:res_requirements} and for high-redshift lightcone construction, as described in \cite{DeRose2018} and \citet{DeRose2021b}.
These simulations were run as part of the multi-resolution "Chinchilla" Simulation suite; the higher-resolution simulations were first used in \cite{Mao2015} and \cite{Lehmann2017}. When presented as "observed" catalogs, these simulations have been referred to as the "Buzzard" simulations.

\begin{deluxetable*}{cccccccc}
  \tabletypesize{\footnotesize}
  \tablecaption{
  Description of simulations used for training and lightcone construction. 
  In this work L2 is mainly used for resolution tests, and L3 is only used for high 
  redshift lightcone construction. Columns describe the simulation name, the minimum
  and maximum redshifts spanned by that simulation ($z_{\rm min}$ and $z_{\rm max}$),
  the periodic box size used to generate the lightcones ($L_{\rm box}$), the number 
  of particles used in each simulation and the particle mass ($N_{\rm part}$ and $m_{\rm part}$),
  as well as the force softening length ($\epsilon_{\rm Plummer}$) and minimum halo mass that central galaxies 
  are populated in ($M_{\rm halo, min}$). \label{table:simulations}}
  \startdata
  \\
  \hline
  Name & $z_{\rm min}$ & $z_{\rm max}$ & $L_{\rm box}$ & $N_{\rm part}$ & $m_{\rm part}$
  & $\epsilon_{\rm Plummer}$
  & $M_{\rm halo, min}$ \\
  \hline
  \hline
  T1 & training only & training only & 400 \hmpc & 2048$^3$& 4.8 $\times 10^{8} \hinv\msun $& 5.5 \hkpc & -- \\
  L1 & 0.0 & 0.32 & 1.05 \hGpc & 1400$^3$ & 3.3$\times 10^{10} \hinv\msun $& 20 \hkpc &  $6\times10^{12}\hmsun$ \\
  L2 & 0.32 & 0.84 & 2.6 \hGpc &  2048$^3$ & 1.6$\times 10^{11} \hinv\msun $ & 35 \hkpc &  $6\times10^{12}\hmsun$\\
  L3 & 0.84 & 2.35 & 4.0 \hGpc & 2048$^3$ & 5.9$\times 10^{11} \hinv\msun$ & 53 \hkpc &  $10^{13}\hmsun $ \\
  \enddata
\end{deluxetable*}

\subsection{Halo Finding}
We identify halos with the publicly available adaptive phase-space
halo finder {\sc Rockstar}\footnote{https://bitbucket.org/gfcstanford/rockstar/}
\citep{behroozi_etal:13a}.  {\sc Rockstar} is highly efficient, and has
excellent accuracy (see for example, the halo finder comparison in
\citealt{Knebe11}). It is particularly robust in galaxy mergers,
important for the massive end of the halo mass function, and in tracking
substructure, important for the abundance matching procedure applied to
\textsc{T1}. We use $M_{\rm vir}$
strict spherical overdensity (SO) masses \citep{bryan_norman:98} here;
additional halo mass definitions are output by {\sc Rockstar} using
these halo centers. {\sc Rockstar} also outputs several other halo
properties, including concentration, shape, and angular momentum
\citep[see][for details]{behroozi_etal:13a}.

\subsection{Merger Trees}

For the highest resolution \textsc{T1} simulation, we track the formation of halos using 100 saved snapshots between $z = 12$ and $z = 0$, equally spaced in $\ln a$.  The gravitationally consistent merger tree algorithm\footnote{https://bitbucket.org/pbehroozi/consistent-trees} 
described in \cite{behroozi_etal:13b} is applied to construct halo merger trees. This algorithm explicitly checks for consistency in the gravitational evolution of dark matter halos between time steps, and leads to robust tracking. Details of the implementation and robustness tests can be found in \cite{behroozi_etal:13b}. Using the resulting merger trees, we are able to track the peak virial mass, $M_{peak}$, and velocity, ${\rm v}_{max}$ value for each identified subhalo.

\subsection{Lagrangian Density Estimation}
\label{subsec:density_estimation}
The final post-processing step for the dark matter simulations before
we can run \addgals\ is to calculate the distance to the $n$-th
nearest particle for both identified halos and all simulation particles.
\addgals\ uses the relation $P(R_{\delta}|M_{r},z)$, where
$R_{\delta}$ is the distance to $n$-th nearest particle, and $n$
is the number of particles whose mass sums to $M_{\delta} = 1.8 \times
10^{13}\hinv\msol$. We measure this Lagrangian density, $R_{\delta}$, for every
particle and halo in the each of the simulations presented in this work. 

\section{Connecting Galaxies to the Matter Distribution}\label{sec:addlums}

We start by describing the first part of the \addgals\ algorithm, which populates a dark-matter-only simulation with galaxies using a model trained on a higher-resolution simulated galaxy catalog. The algorithm is designed to work on the matter distribution from either a simulation snapshot or lightcone output. A key strength of the algorithm is its ability to use relatively low-resolution dark matter
simulations. Consequently, we operate directly on the dark matter
particle distribution, using the density information described in 
\cref{subsec:density_estimation} to assign galaxy properties.
The algorithm is designed to insert galaxies with 
single-band absolute magnitudes. While this quantity could be chosen to be 
anything that is reasonably well-correlated with density, in the present 
work we use the SDSS $r$-band magnitude $k$-corrected to $z=0.1$, $M_{\rm r}^{0.1}$, hereafter $M_{\rm r}$.

Here we train \addgals\ to reproduce the galaxy--dark matter connection in an  abundance matching (SHAM) model. We use the best-fit model from \cite{Lehmann2017} to assign galaxies to dark matter halos.  This procedure, including the luminosity function and implementation details, are described in \cref{sec:shamcat}.
In principle, the same type of mapping can be tuned to other catalogs, such those constructed with SAMs, hydrodynamical simulations, or other empirical models.
\addgals\ is able to approximate catalogs produced by the SHAM model because 
local density measurements contain information about halo mass and halo-centric
distance, allowing for the accurate reproduction of the HOD and radial profiles of
the catalog that \addgals\ is tuned to. Limitations of the SHAM catalog that 
\addgals\ is tuned to, such as the effects of artificial subhalo disruption on the
satellite populations of massive halos, are also inherited.

Broadly, this part of the \addgals\ algorithm proceeds in two steps, described in the following subsections:
\begin{enumerate}
\item Central galaxies are placed on all resolved host halos above the some minimum halo mass threshold, $M_{\rm min}$, as listed in \cref{table:simulations} (\cref{sec:cenpop}).
\item The one-dimensional probability density function (PDF) of local dark matter density 
around galaxies conditioned on absolute magnitude measured from the SHAM model is used to assign the rest of the galaxies (\cref{sec:unrespop}). 
\end{enumerate}

\subsection{Populating Resolved Central Galaxies}
\label{sec:cenpop}

A statistical relationship between halo mass its primary galaxy's absolute magnitude, $p(M_{\rm r,cen} | M_{\rm vir})$,
is assumed in order to populate central galaxies. The mean of this distribution is given by
\begin{equation}
\langle M_{\rm r,cen} \rangle  (M_{\rm vir}) =  M_{\rm r,0} - 2.5 [ a\log x - (1/k)\log(1+x^{bk}) ],
\label{eq:bcg_ml}    
\end{equation}
where $x = M_{\rm vir}/M_{\rm c}$ and $a, b, k, M_c$, and $M_{r,0}$ are redshift-dependent fitting parameters. This form was proposed by \cite{vale_etal:06} to match early SHAM catalogs and has 
been shown to provide a good fit to observational catalogs \citep{hansen_etal:09, zheng_etal:07}. 
A Gaussian scatter in absolute magnitude at fixed mass is assumed such that 
a halo with mass $M_{\rm vir}$ is assigned a magnitude drawn from
\begin{equation}
p(M_{\rm r,cen} | M_{\rm vir}) = \mathcal{N}(\langle M_{\rm r,cen}\rangle (M_{\rm vir}), \sigma_{M_{\rm r,cen}})
\label{eq:pcen}
\end{equation}
where $\sigma_{M_{\rm r,cen}} = 0.425$, matching the scatter assumed in the SHAM model.
Tests have shown that this relation must be applied at least for all host halos more massive 
than $M_{\rm vir} \sim 10^{13}\, \hMsun$ to accurately reproduce the projected 
clustering of luminosity-selected galaxies (see \cref{sec:res_requirements} for further discussion of resolution requirements).

\Cref{eq:bcg_ml} is then fit to the SHAM catalog in each time snapshot over the mass 
range $10^{12}\, \hMsun \le M_{vir} < 10^{15}\, \hMsun$. When populating lightcone simulations, the fit from the snapshot that is closest to the redshift under consideration is used. Evolution in this relation over the redshift ranges between snapshots is negligible. Validation of this relation is described in \cref{app:more_validation}. Once $p(M_{\rm r,cen} | M_{\rm vir})$ has been determined, it is used to populate all central galaxies down to the halo mass limits in \cref{table:simulations} by sampling from the distribution in \cref{eq:pcen}, conditioned on the mass of each host halo in the simulation.  We note that the resolution of L1 would enable us to go to lower masses, but this is not required to match the clustering properties of the higher-resolution simulation (\cref{sec:res_requirements}), so we keep this limit constant between L1 and L2 for continuity.

\subsection{Populating Galaxies in Unresolved Structures}
\label{sec:unrespop}
The resolution of the lightcone simulations used here is such that 
central galaxies assigned using the method described above constitute only a small fraction of all galaxies that would be observed by deep photometric surveys. To populate the rest of the galaxies, the relationship between large-scale dark matter density and galaxy rest-frame magnitude, $p(R_{\delta}|M_r,z)$, is used. $\rd$ is defined as the radius enclosing a mass scale of $M =1.8\times10^{13}\, \hmsun$, characterizing the local dark matter density around galaxies. For the \textsc{L1} simulation, this radius corresponds to the distance to the 538th nearest dark matter particle. \Cref{sec:pop} describes how this relation is determined from a SHAM catalog. This mass scale is roughly equivalent to $M_*$, the typical collapsing halo mass, at $z=0$ for this cosmology, and thus effectively distinguishes between halos of different biases.

In order to parallelize the \addgals\ algorithm, we divide the lightcone simulations into domains of approximately
$(200\, \hMpc)^3$ in volume. This is accomplished by dividing the sky in angle as well as redshift. In a given patch with a redshift range of $z_{\mathrm{low}} < z < z_{\mathrm{high}}$,
we create a catalog of galaxies with magnitudes and redshifts $\{M_{r,i},
z_{i}\}$, where $i=1,\ldots,N$, and $N$ is the total number of galaxies, given by:

\begin{align}
N=&\int_{z_{\mathrm{low}}}^{z_{\mathrm{high}}}dz\frac{dV}{dz}\int_{-\infty}^{M_{r,min}(z)}\phi_{\mathrm{unres}}(M_{r},z),
\end{align}
\noindent
where $M_{\rm r,min}$ is the faintest absolute magnitude that we populate galaxies to, typically chosen to yield a catalog complete to a particular observed $r$-band magnitude limit, and $\phi_{\mathrm{unres}}$ is the luminosity function of all objects to be placed on unresolved structures in the simulation. This function subtracts central galaxies from the total luminosity function in order to avoid double-populating bright galaxies, and thus is specified by 
\begin{align}
\phi_{\mathrm{unres}} (M_{\rm r},z) =& \phi (M_{\rm r},z) - \phi_{\mathrm{res}} (M_{\rm r},z) \nonumber \\
=& \phi (M_{\rm r},z) \nonumber \\
&-\int_{M_{\rm min}}^{\infty}dM_{\mathrm{vir}}p(M_{\rm r,cen}|M_{\mathrm{vir}},z)n(M_{\mathrm{vir}},z),
\end{align}
where $n(M_{\mathrm{vir}},z)$ is the halo mass function in the simulations. We modulate the normalization of the luminosity function by the local dark matter overdensity,
\begin{equation}\label{eq:norm_modulation}
    \phi_{i,local} = \phi_{i} (1 + \delta) .
\end{equation}

Here $\phi_{i,local}$ is the normalization of the luminosity function in the local domain $i$ and $\delta$ is the matter overdensity within the domain. This avoids fixing the number density of galaxies on the scale of the domain size, but can induce a scale-dependent bias on scales of similar size to the domains. We can see the reason for this as follows. Note that $b(r) = \delta_g / \delta_m$, where the overdensities to are smoothed on a radius $r$. \Cref{eq:norm_modulation} enforces $\delta_g(r>r_{\rm domain}) = \delta_m(r_{\rm domain})$, where $\delta_m(r_{\rm domain})$ is the matter overdensity on the scale of the domain size. Thus, $b(r>r_{\rm domain}) = \delta_m(r_{\rm domain}) / \delta_m(r)$, and is scale dependent for $r>r_{\rm domain}$. The domains are taken to be at least $(200 \hMpc)^3$ and thus the effect of this choice is negligible for most applications.  However, this choice may have an impact on covariance matrices involving scales larger than the domain size, which requires further investigation. The size of these domains is chosen as a compromise between this effect and the run time of the algorithm, as \addgals\ can be run in parallel over each domain. 

Redshifts of each galaxy, $z_{i}$, are then drawn from
\begin{equation}
P(z) = \frac{1}{N} \, \frac{dV}{dz} \int_{-\infty}^{M_{r,min}}dM_{r}\phi_{\mathrm{unres}}(M_{r},z), 
\end{equation}
and magnitudes, $M_{\mathrm{r},i}$, are drawn from $\phi_{\mathrm{unres}}(M_{r},z_{i})$.
With magnitudes and redshifts assigned to every galaxy, densities $\{R_{\delta,i}\}$ are drawn from $p(R_{\delta}|M_{r,i},z_{i})$, and each galaxy is then assigned to
a particle, going from brightest to faintest, with the closest match in redshift and 
density that has not already been assigned. The details of this process are described 
in \cref{app:density-sampling}.

At this point, the described algorithm produces a catalog of galaxies with
$r$-band absolute magnitudes. This algorithm is a very efficient way to generate a 
large-volume synthetic catalog with faint galaxies using primarily simulations with 
modest resolution. Additionally, as long as abundance matching in the $r$-band 
works well at high redshifts, we expect that the galaxy distribution 
should match the clustering at a wide range of redshifts. Note that the 
same algorithm can also be used to populate comoving snapshots by fixing $z$ in the above equations to the redshift of the snapshot output. The next 
section describes how $p(R_{\delta}|M_{r,i},z_{i})$ is determined from a SHAM catalog
on a high-resolution simulation.

As we show in \cref{sec:lumtests} and \cref{app:more_validation}, with this algorithm we can 
create a galaxy catalog that matches the projected galaxy two-point function, halo occupation distribution, 
conditional luminosity function, and galaxy profiles in halos of a galaxy catalog populated using SHAM in a higher-resolution simulation.

\subsection{Determining the $p(R_{\delta}|M_r,z)$ relation}
\label{sec:pop}

The form of $p(R_{\delta}|M_r,z)$ is the crux of the \addgals\ algorithm. We have found the following bi-modal form is a good fit to our simulations,
\begin{align}
\label{eq:pdf}
p(R_{\delta}| M_r<x, z) =& p(R_{\delta};\Theta(x,z)) \\
 =& (1-p)e^{-({\rm ln}(R_{\delta})-\mu_c)^2/2\sigma_c^2}/R_{\delta}\sqrt{2\pi} \sigma_c \nonumber \\
 & + pe^{(R_{\delta}-\mu_f)^2/2\sigma_f^2}/\sqrt{2\pi} \sigma_f \nonumber\ .
\end{align}
This gives the probability that a galaxy with magnitude $M_r<x$ at 
redshift $z$ has a local dark matter density, $R_{\delta}$. 
Each of this relation's five free parameters, 
$\Theta(x,z) = \{\mu_c, \sigma_c, \mu_f, \sigma_f, p\}$, are functions of galaxy 
absolute magnitude and redshift, and the dependence of $\Theta(x,z)$ on these variables 
is modeled using a Gaussian process as described in \cref{app:density-modeling}.

\begin{figure*}
\centering
\includegraphics[width=\columnwidth]{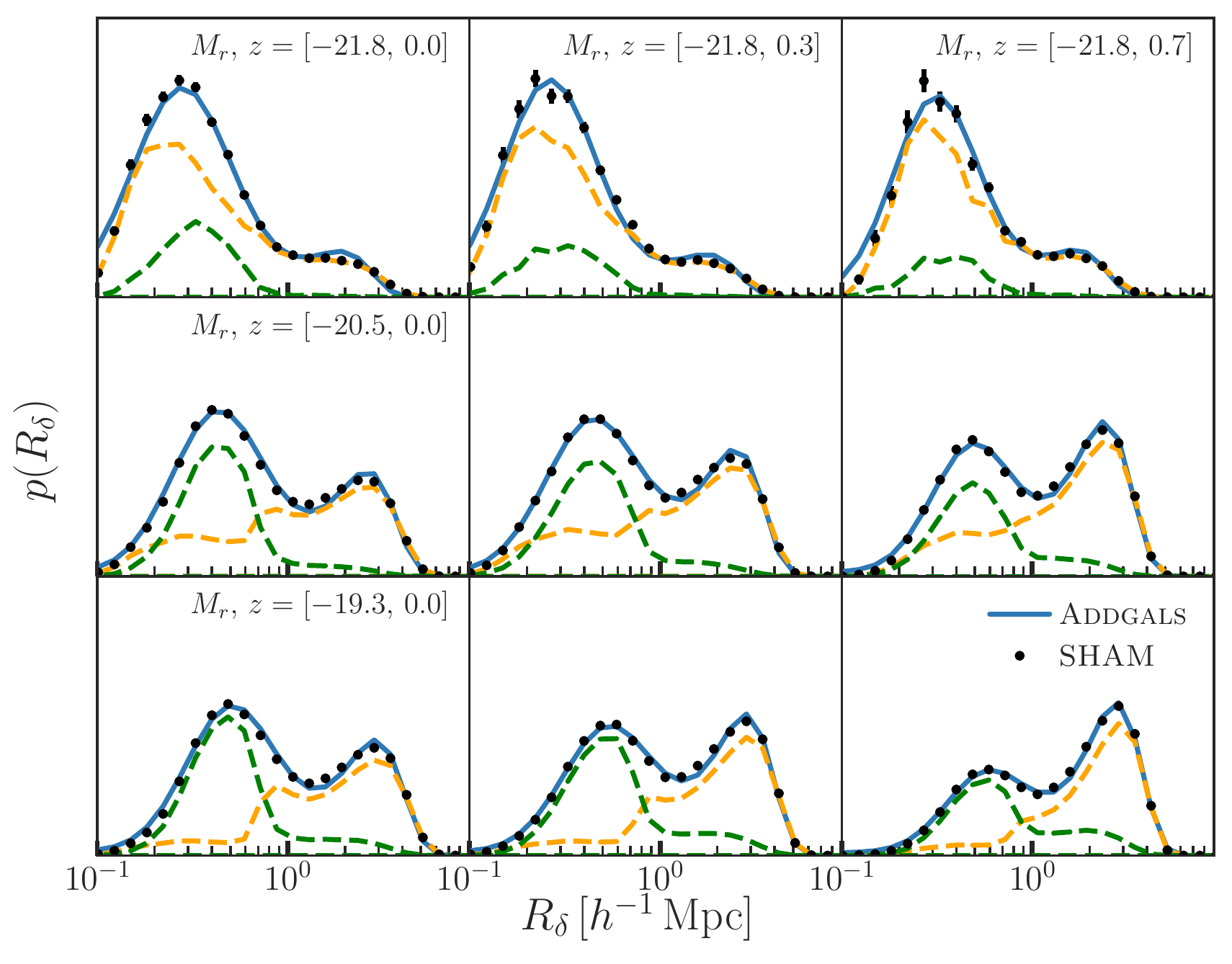}
\includegraphics[width=\columnwidth]{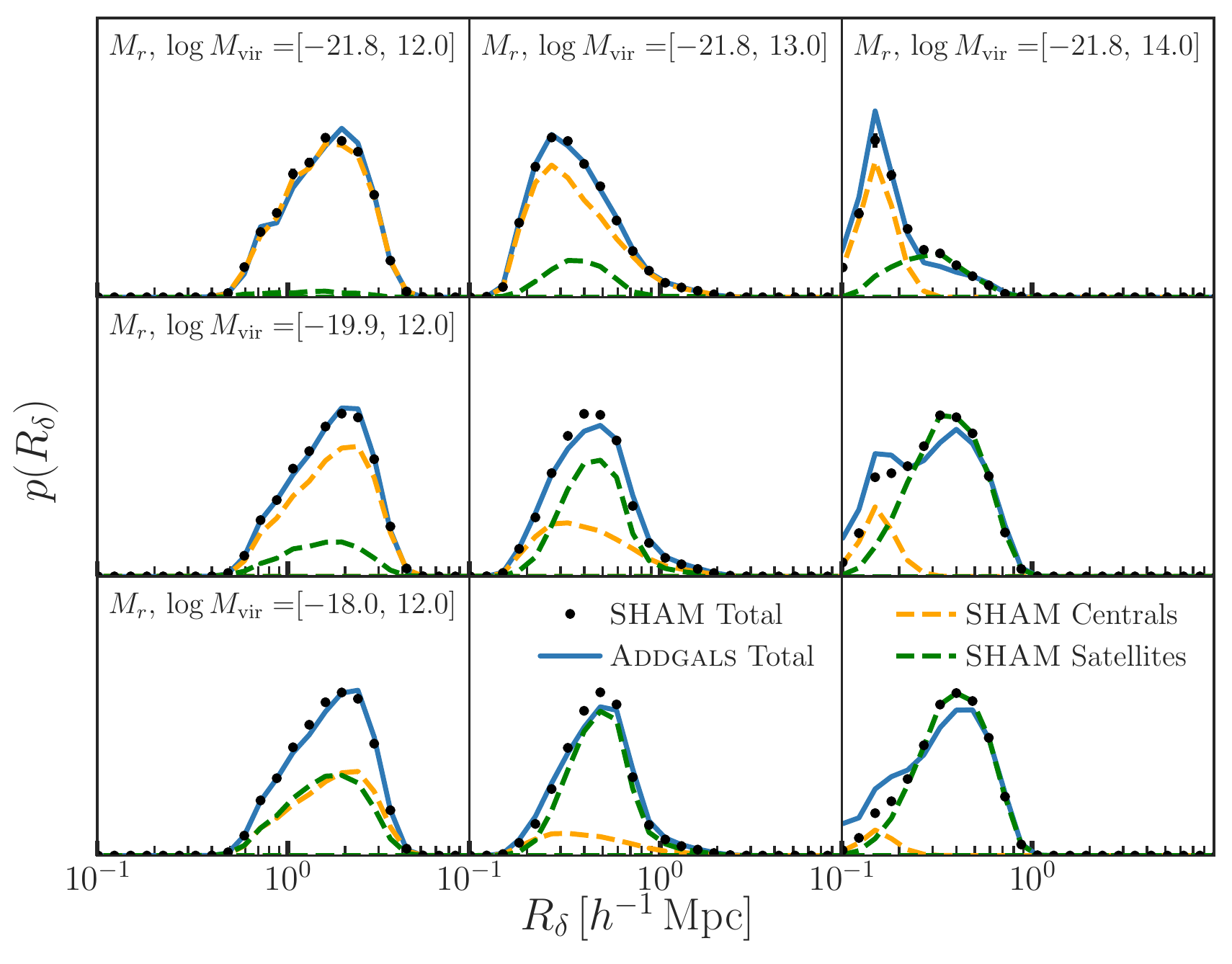}
\caption{
PDF of dark matter densities, characterized by \rd, the radius enclosing a mass of $1.8 \times 10^{13}\hMsun$, for various galaxy populations in the simulations. Both panels compare the \textsc{Addgals} distribution (blue lines) to the SHAM distribution (black points) in the T1 simulation.  Dashed green and orange lines represent the $\rd$ distributions for satellite and central galaxies, respectively, in the SHAM catalog.
\emph{Left panel} compares the \rd\ distributions for galaxies, binned by absolute magnitude (rows),residing in all halo masses at different redshifts (columns).  Fainter central galaxies tend to live in less massive halos, corresponding to large \rd, while satellites are hosted by more massive halos corresponding to small \rd, leading to the observed bi-modality in $p(\rd)$. \emph{Right panel} compares the 
 \rd\ distributions at  $z=0$ for galaxies as a function of absolute magnitude (rows) and host halo mass (columns). The agreement shows that \textsc{Addgals} is able to successfully reproduce the density distribution of the SHAM model. 
}
\label{fig:allpdfs}
\end{figure*}

Figure \ref{fig:allpdfs} shows the distribution of $p(\rd)$ for bins in galaxy magnitude and redshift (left) and galaxy magnitude and host halo mass (right).  The full distribution $p(\rd | M_r<x, z)$ in bins of magnitude and redshift in the input SHAM model (points) is well reproduced by the \addgals\ model applied to the T1 simulation (blue line). The reduced chi-squared values for these fits can be large, $O(10-100)$, but the median absolute deviation is less than $2.5\%$ for all redshift and magnitude bins.

The orange and green lines in this figure show the same distributions for central and satellite galaxies in the SHAM, to indicate which region of this distribution they populate. At low luminosity (bottom rows), these two populations are easily separated by density. At higher luminosity (top row), the two populations cannot be distinguished by density; this motivates separate modeling of bright central galaxies through $\langle M_{\rm r,cen} \rangle  (M_{\rm vir})$ so that central galaxies can be distinguished from bright satellites in massive systems.

The right side of \cref{fig:allpdfs} shows the distribution of $p(\rd)$ in bins of halo mass and galaxy magnitude at redshift $z=0$, $p(\rd | M_{\rm vir}, M_r<x, z=0)$.  Splitting the distribution in this way gives intuition for how assigning galaxies by \rd\ can approximate assignment by halo mass, even in simulations that are lower resolution than the relevant resolved halos. At high mass (right column), we see that this distribution is highly peaked towards small \rd (high densities), and satellites and centrals are easily separated in \rd\ space.
The movement of this peak with mass is what enables assignment by \rd\ to distinguish between different halo masses. The smoothing scale used here, ($1.8\times 10^{13}\, \hmsun$), effectively distinguishes more biased halos above the smoothing scale from lower mass halos below the smoothing scale (left most column), where halo bias is relatively flat. For halos below the the mass smoothing scale, the distribution $p(\rd | M_{\rm vir}, M_r<x, z)$ broadens, and  there is much more scatter in $M_{\rm vir}$ when assigning by \rd. Due to this broadening, the \addgals\ algorithm is susceptible to scattering galaxies between different halo masses at fixed \rd. This can lead to Eddington-like biases in halo occupation distributions in \addgals, where galaxies that should be placed in halos with masses less than the smoothing mass scatter up into halos with masses approximately at the smoothing mass. However, in this regime, halo bias is relatively flat so this scatter does not significantly impact the projected clustering signals in \addgals. 

\subsection{Algorithm Overview}\label{sec:overview}

The algorithm steps can be summarized as follows:

\begin{enumerate}
\item High-resolution modeling: apply a SHAM model to high-resolution $N$-body snapshots
\item High-resolution training:
\begin{enumerate}
\item calibrate the luminosity--density--redshift relation, $p(R_{\delta} |M_r,z)$ from the SHAM model 
\item calibrate the central luminosity--halo mass relation, $p(M_{\rm r,cen} | M_{\rm vir})$ from the SHAM model
\end{enumerate}
\item Populating lightcone simulations:
\begin{enumerate}
\item populate central galaxies based on $p(M_{\rm r,cen} | M_{\rm vir})$
\item populate the rest of the galaxies based on $p(R_{\delta} |M_r,z)$ and the luminosity function
\end{enumerate}
\end{enumerate}

The final result is a synthetic galaxy catalog containing positions, velocities, 
and single-band photometric information. Next we validate the steps above using 
observations from the SDSS.

\begin{figure*}
\centering 
\includegraphics[width=1.1\columnwidth]{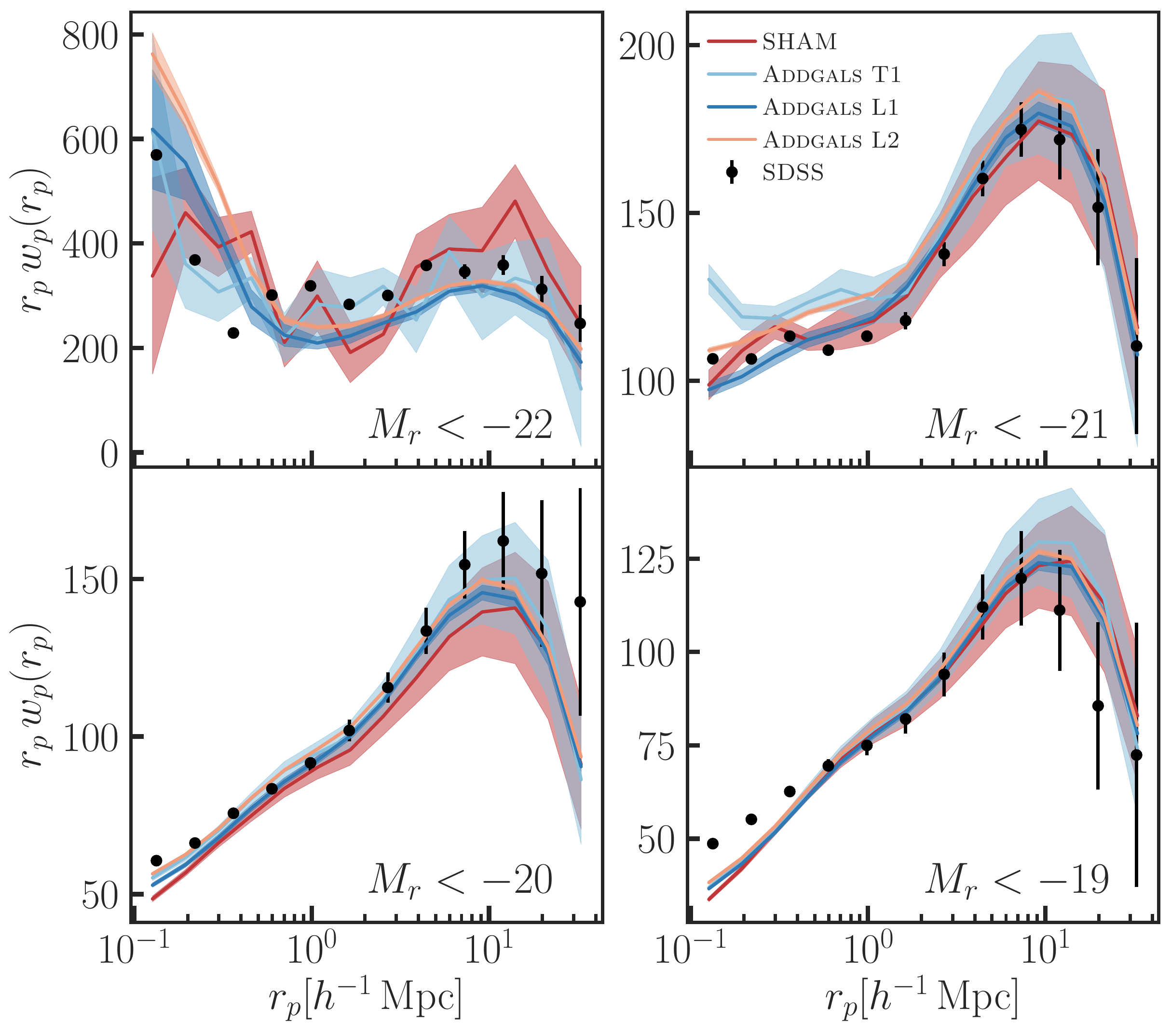}
\includegraphics[width=\columnwidth]{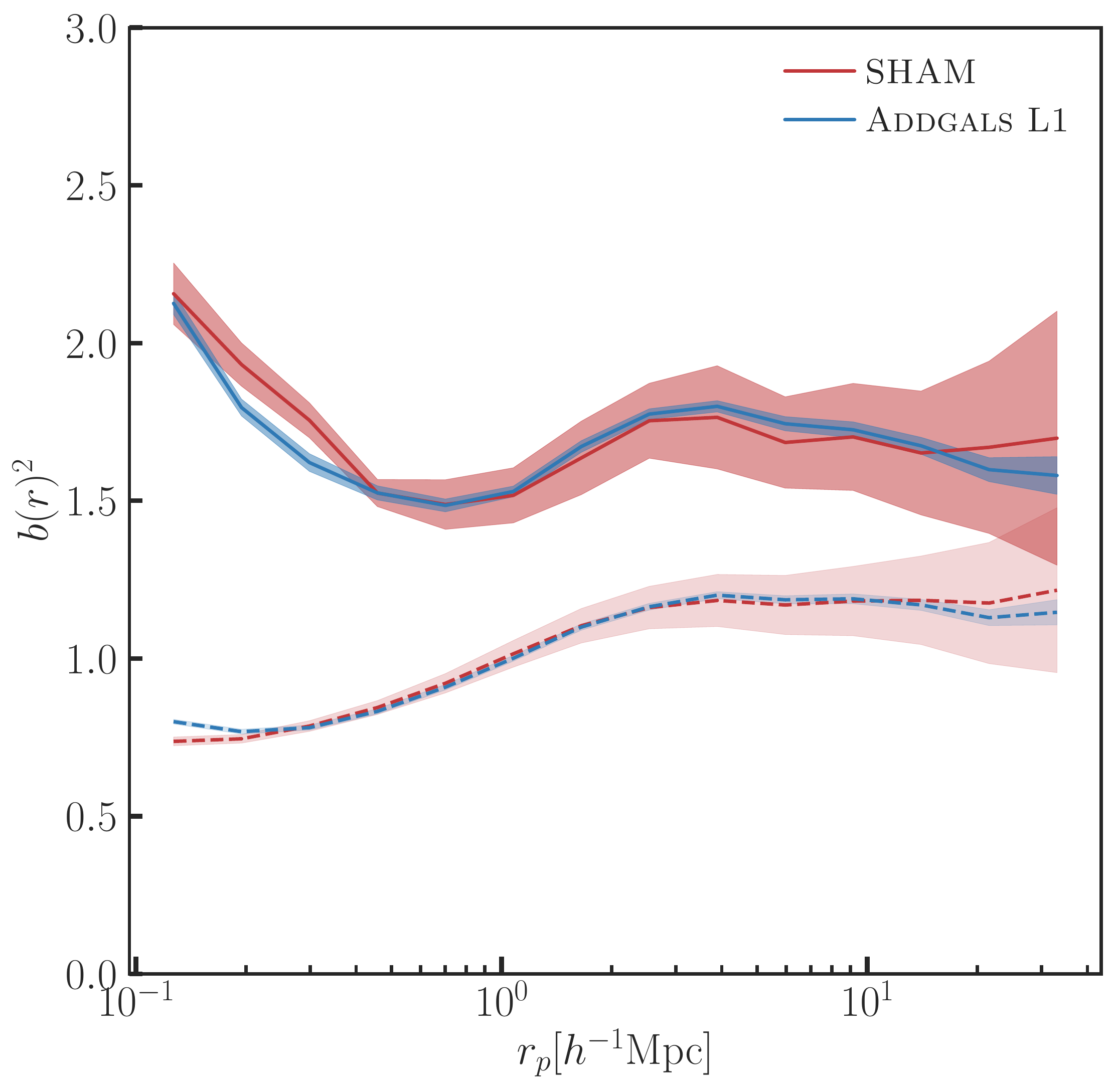}
\caption{Comparison of \addgals\ clustering and bias with that measured in the SHAM model it is tuned to.
\emph{Left}: Projected correlation functions for a SHAM model applied to the $z=0$ snapshot of the \textsc{T1} simulation, and \textsc{Addgals} models trained on that SHAM run on the $z=0$ snapshots of the \textsc{T1}, \textsc{L1}, and \textsc{L2} simulations. Shaded regions and error bars are the $1\sigma$ errors, estimated via a jackknife procedure described in the text. Each panel shows a different magnitude threshold.
The corresponding SDSS measurements from \citet{Reddick12} are shown for comparison.
Agreement between \addgals\ catalogs based on different resolution simulations and the SHAM model are generally good. Discrepancies between the small-scale simulation measurements and data are discussed in \cref{sec:lumtests}. \emph{Right}: Bias as a function of scale for the \textsc{L1} \addgals\ catalogs and the SHAM model it is tuned to, indicating good agreement. Solid lines are for all galaxies with $M_r<-21$ and dashed lines are $M_r<-19$.}
\label{fig:wpcomp}
\end{figure*}

\section{Validation of the Luminosity--Density Assignment}
\label{sec:lumtests}

Here, we present a number of tests validating the ability of \addgals\ to reproduce the properties of the \textsc{T1 SHAM} model in the lower-resolution \textsc{L1} simulation. The tests in this section compare an \addgals\ catalog run on the $z=0$ snapshot output of the \textsc{L1} simulation and a \textsc{SHAM} catalog run on the $z=0$ snapshot output of the \textsc{T1} simulation unless otherwise noted.

The left side of \cref{fig:wpcomp} compares the projected correlation function of the \textsc{T1 Addgals}, \textsc{L1 Addgals}, and \textsc{L2 Addgals} catalogs with the \textsc{T1 SHAM} catalog and the SDSS measurements presented in \citet{Reddick12}. The function $w_p(r_{p},\pi)$ is measured in the snapshots using the Landy--Szalay \citep{landy_szalay:92} estimator, i.e. $w_{p}(r_{p},\pi) = (DD - 2DR + RR)/RR$, using 13 logarithmically spaced bins in $r_p$ between $0.1\hmpc$ and $40\hmpc$, subsequently integrating these measurements along the line-of-sight out to  $\pi_{max}=60\,\hmpc$ to obtain $w_p(r_p)$. Ten times as many random points as galaxies are used to estimate $DR$ and $RR$, where the randoms are distributed uniformly in each sub-volume. Errors are estimated via jackknife using 64 sub-volumes for each simulation.

We use the best-fit \textsc{SHAM} model from \citet{Lehmann2017}, and as such the agreement between the \textsc{SHAM} catalog and SDSS is good, albeit with relatively large errors on the \textsc{SHAM} measurements for the brighter samples. Details of how the \textsc{SHAM} catalog are constructed are described in \cref{sec:shamcat}. The \addgals\ catalog and the \textsc{SHAM} catalog are consistent with each other at most scales and magnitudes. Discrepancies between the \addgals\ catalogs and the SDSS data can be seen in the $M_{r}<-22$ and $M_{r}<-19$ measurements. Given the large errors on the \textsc{SHAM} catalog for the $M_{r}<-22$ magnitude cut, it is unclear whether this discrepancy is due to a disagreement between \addgals\ and the \textsc{SHAM} model, or whether \addgals\ and the \textsc{SHAM} model agree well, and the SHAM model disagrees with the data. \citet{Lehmann2017} finds a marginal preference for lower scatter at brighter luminosities, which is consistent with the latter of these two possibilities. We also make use of a slightly different luminosity function than \citet{Lehmann2017}, with the main difference coming at the brightest end, where our luminosity has a shallower slope. This may also lead to a reduced clustering amplitude in the $M_r\le -22$ bin.

For $M_{r}<-19$, \addgals\ and \textsc{SHAM} catalogs are in good statistical agreement for $r_p>1\,\hmpc$, but deviate significantly from SDSS at scales $r_p<1\,\hmpc$. The \textsc{SHAM} model suffers from artificial subhalo disruption in this regime, leading to lower small-scale clustering, and the \addgals\ model has inherited this issue through the $p(\rd | M_{r},z)$ distribution. Differences between the \addgals\ catalogs due to differences in simulation resolution are discussed in \cref{sec:res_requirements}.

The right hand side of \cref{fig:wpcomp} compares the behavior of galaxy bias between the \addgals\ and \textsc{SHAM} catalogs. The bias measurements are made by taking the ratio of $w_p(r_p)$ for galaxies and that measured on the matter distribution in each respective simulation. Given the agreement between the projected correlation functions of the \textsc{SHAM} model and \addgals\, the agreement seen here is expected. A notable feature in this figure is the scale at which the different samples conform to a linear bias model, i.e. $\delta_{g}(r) = b_1 \delta_{m}(r)$, on large scales. For the fainter $M_{r}<-19$ sample, galaxy bias becomes linear in both catalogs for scales with $r_{p}>4 \hmpc.$ For the brighter $M_{r}<-21$ sample, the \textsc{SHAM} catalog also appears to behave linearly for $r_{p}>4 \hmpc$. The \addgals\ measurements are fully consistent with the noisier SHAM measurements, but the smaller error bars in these measurements show hints of non-linear bias out to slightly larger scales. This result is expected for this more massive galaxy sample.

\begin{figure*}
\centering
\includegraphics[width=\linewidth]{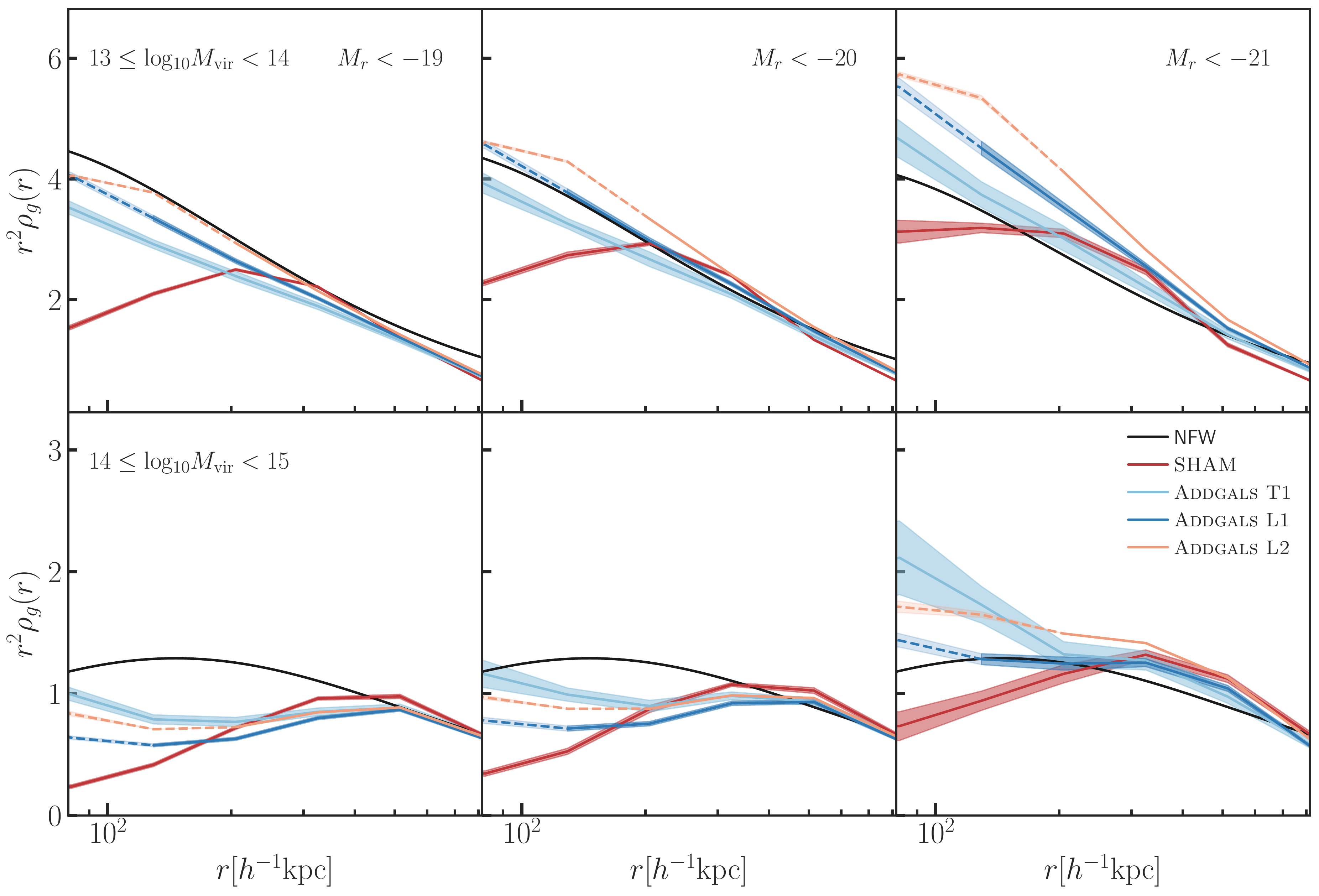}
\caption{
Radial profiles of galaxies in group- and cluster-sized halos 
for SHAM, \textsc{Addgals T1}, \textsc{Addgals L1}, and \textsc{Addgals L2} catalogs. Columns show different absolute magnitude cuts given by the labels in each panel, and rows show different halo mass bins. Each panel is normalized such that the \textsc{SHAM} curves pass through 1 on the largest scale plotted; uncertainties for each measurement are estimated using jackknife. The lines transition from solid to dashed at five times the force softening length of each respective simulation in order to approximate the scale where we expect resolution to affect the matter profiles of these halos. The black line indicates  an NFW profile for the mean halo mass in the bin, using the \citet{Diemer2019} mass--concentration model and normalized to match the SHAM profiles at $R_{\rm vir}$. Agreement between the catalogs is generally good at scales larger than $200\hkpc,$ which approximately corresponds to the scale imposed by the mass used to calculate $R_{\delta}$. This scale is relatively independent of halo mass, modulo changes in the mean halo concentration as a function of halo mass. Below this scale, $R_{\delta}$ becomes approximately constant and the \textsc{SHAM} and \addgals\ profiles no longer track each other. At scales less than this smoothing scale in the less massive bin the \addgals\ profiles approximate NFW profiles much more closely than the SHAM catalog, which is affected by artificial subhalo disruption on these scales. On the smallest scales depicted here, the \addgals\ profiles begin to deviate near the resolution limits of each respective simulation, with the slope of the profiles turning over in a characteristic manner. In the more massive bin, none of the catalogs is well described by an NFW profile. At these masses, the \addgals\ smoothing scale can probe smaller fractions of the halo virial radius, and subhalo disruption effects that become important for satellite galaxies in the \textsc{SHAM} are inherited by $p(\rd | M_r)$, and thus the \addgals\ catalogs. This causes deviations from NFW profiles for all catalogs above the \addgals\ smoothing scale. For a more quantitative discussion, see \cref{sec:lumtests}.}
\label{fig:rprof}
\end{figure*}

We compare the radial profiles of galaxies around host halos between the \addgals\ and \textsc{SHAM} catalogs in \cref{fig:rprof}. The measurements from \addgals\ run on both the \textsc{T1} (light blue) and \textsc{L1} (dark blue) and \textsc{L2} (orange) simulations are included. All curves are normalized so that the SHAM radial profile equals one on the largest scale in the figure. We indicate where we expect resolution effects in the matter density profiles of the host halos by plotting curves below this scale with a dashed line. This scale is approximated by five times the force softening scale used in each simulation.

We see that these radial profile measurements exhibit significant differences between \addgals\ and \textsc{SHAM}. Above $\sim 200\,\hkpc$ the two models agree well, as expected by the agreement between projected correlation functions for the two models. Below this scale in both mass bins shown here the \textsc{SHAM} catalog experiences a flattening, and becomes inconsistent with the expectation of profiles with Navarro-Frenk-White (NFW) functional forms shown by the black line \citep{NFW}. The NFW prediction shown here assumes the mean host mass in the bin, the mass--concentration relation from \citet{Diemer2019}, and is normalized so that it matches the SHAM curves at $R_{\rm vir}$.  The reason for this deviation from an NFW expectation for \textsc{SHAM} is likely artificial subhalo disruption for halos which have close pericentric passages \citep{vandenbosch2018a,vandenbosch2018b}. 

We can understand the behavior of the \addgals\ measurements in the following way. Under the assumptions of the \addgals\ algorithm, it is possible to write the radial profile of galaxies of absolute magnitude $M_r$ in a halo of mass $M_h$ as
\begin{align}
    \rho(r | M_r, M_h) &\propto \int p(r| \rd, M_h) p(\rd | M_r) d\rd \\
                        &= \int \frac{p(r | M_h) p(\rd | r, M_h)}{p(\rd | M_h)} p(\rd | M_r) d\rd \, .
\end{align}
$p(r| \rd, M_h)$ is the PDF of matter particle distances to centers of halos of mass $M_h$, given that the particle has a local density measurement of \rd. The first line above follows from the fact that \addgals\ assigns a galaxy with absolute magnitude $M_r$ to a random particle with density $\rd$, where $\rd$ is a random draw from $p(\rd | M_r)$. The second line above is obtained from the first via application of Bayes theorem. 

We see that we can express \addgals\ radial profiles in terms of $p(\rd | M_r)$, the normalized radial profile of matter in halos of mass $M_h$, $p(r | M_h)$, the distribution of \rd\ as a function of halo-centric radius, $r$, $p(\rd | r, M_h)$, and $p(\rd | M_h)$. Note that in the limit where $p(\rd | r, M_h)$ is constant as a function of $r$, then $\rho(r | M_r, M_h) \propto p(r | M_h)$. This is the case on small scales ($\lesssim 200 \hkpc$), where $M^{*}$, the mass scale used to calculate \rd, is significantly larger than the mass enclosed within radius $r$. This makes it clear that the flattening of the slope in the \addgals\ \textsc{L1} and \textsc{L2} catalogs on small scales in the lower mass bin of \cref{fig:rprof} is due to a flattening in the actual matter profiles in halos at those scales, which is close to five times the force softening radii used in these simulations where such a turn over is expected \citep{DeRose2018a}. When using a higher-resolution simulation like \textsc{T1}, this flattening is no longer seen. 

For the lower mass bin, this implicit smoothing scale is approximately the same size as \rvir, and so the \addgals\ profiles approximate an NFW profile well for the entire one-halo term. For the higher mass bin, the smoothing scale is significantly less than \rvir\ and so the one-halo term exhibits significantly more complicated behavior. Above the smoothing scale, the \addgals\ profiles track the \textsc{SHAM} catalog profiles well. At scales less than \rvir\ but greater than the smoothing scale, the \textsc{SHAM} catalog is significantly flattened by artificial subhalo disruption, and so on these scales the \addgals\ catalog inherits the same flattened profile. On scales below the smoothing scale, the \addgals\ catalog reverts to being proportional to the matter profile, leading to the observed upturn relative to the SHAM catalog. This flattening of the \addgals\ profiles at high mass has significant implications for optical galaxy cluster finding, as it leads to a deficit of galaxy number densities in massive clusters with respect to that observed in data as discussed in \cref{sec:color-env}. Additional validation of the \addgals\ catalogs is provided in \cref{app:more_validation}.

We note that a precise match between $P(R_{\delta}|M_{r},z)$ in \addgals\ and \textsc{SHAM} is extremely important in achieving the level of agreement seen in the above comparisons. Even relatively small changes in this distribution make these matches appreciably worse. These comparisons demonstrate the real strength of the \addgals\ algorithm.  Here, we are able produce simulated galaxy catalogs that are complete in absolute magnitude and reproduce observed galaxy clustering properties using an $N$-body simulation that has a particle number density only a factor of $\sim 100$ higher than the galaxy density.  Additionally, tests have shown that such an agreement is possible for simulations with significantly lower resolution, where the number densities differ by only as much as $\sim 20$. The ability to create a realistic galaxy distribution on such modest resolution simulations allows the creation of very large volume synthetic catalogs, appropriate for modeling large photometric and spectroscopic surveys such as SDSS, DES, and LSST, without resorting to any sort of replication techniques or highly expensive simulations, as would be required for most other algorithms.

\section{Assignment of Galaxy SEDs}
\label{sec:addcols}

Once the galaxies have been populated with phase-space positions and $r-$band luminosities, we assign SEDs to each galaxy in the second part of the \addgals\ algorithm. While the SED assignment algorithm was developed in conjunction with the galaxy assignment method discussed above, the algorithm is independent and able to operate on any galaxy catalog that already has absolute magnitudes defined in one band. This part of the algorithm, which is referred to as \textsc{Addseds} (Adding Density-Determined SEDs) when used on its own, has been used independently from the first step of \textsc{Addgals} in previous works, based on earlier versions of the present algorithm \citep[see, e.g.,][]{gerke_etal:12, Mao2018}.

\textsc{Addseds} assumes that galaxy SEDs are set by both absolute magnitude and galaxy environment and uses a training set consisting of the SDSS DR7 VAGC \citep{blanton05}, whose SEDs are mapped onto the simulated galaxies.  We cut the training set to $0.005 < z < 0.2$, since the bright, higher redshift objects and very faint low redshift objects represent a biased sample with respect to the rest of the population of galaxies. The final training set consists of approximately 600,000 spectroscopic galaxies from the SDSS main sample.

For each galaxy in the simulated galaxy catalog, the set of SDSS galaxies in a bin of $M_{r}$ around each simulated galaxy are identified and one SED is randomly chosen from this set and assigned to the simulated galaxy. The assumed bin width depends on $M_r$ as $\Delta M_r = \Delta M_{r,0} (22.5 + M_r)$, where $\Delta M_{r,0}=0.1$. If no galaxy in SDSS is found, the bin width is relaxed to $\Delta M_r = \Delta M_{r,0} (22.5 + M_r)^2$; the latter criteria always enables a match to be found. These bin widths have been tuned order to minimize discreteness effects in color space due to assigning the same SDSS SED repeatedly to many simulated galaxies. 

After this initial step, an environmental dependence of the SED assignment is imparted on the catalog. This is accomplished by correlating rest frame $g-r$ color with a local density proxy that can be accurately measured in modest resolution \nbody\ simulations. The proxy that we have found to work well is the distance between the galaxy in question and the nearest halo above a mass threshold, $M_{\textrm{cut}}$. We refer to this proxy as $R_{h}$. 

In detail, $g-r$ colors are mapped onto the simulated galaxies by enforcing the following ansatz:
\begin{align}
\label{eq:cam}
    P(< g-r | M_r) = P(<\widetilde{R_{h}} | M_r)
\end{align}

\noindent where $\widetilde{R_h}$ is a noisy version of $R_h$, and the Pearson correlation coefficient between $Rank(R_h)$ and $Rank(\widetilde{R_h})$ is set to $r_{\rm corr}$. In doing so, we allow for an imperfect correlation between $R_h$ and $g-r$, which is necessary to match observed clustering statistics as a function of $g-r$. In order to reduce discreteness effects, $P(< g-r | M_r)$ and $P(<\widetilde{R_{h}} | M_r)$ are computed in sliding windows around each galaxy, such that the width of the window in $M_r$ yields 100 galaxies with which to estimate the above distributions. The values for $M_{h, \textrm{cut}}$ and $r_{\rm corr}$ are free parameters of the model, tuned to reproduce color dependent clustering in SDSS. This work uses the best fit values of these parameters from \citet{DeRose2021}, where additional implementation details this model can be found.

This is very similar in spirit to Conditional Abundance Matching (CAM) models that first assign magnitudes to halos via abundance matching, and then assign colors to galaxies at fixed magnitude by making the ansatz that $Rank(X_{h}) \sim Rank(g-r)$, where $X_{h}$ is usually taken to be a dark matter halo property such as formation time or accretion rate \citep{Masaki2013, Hearin2014, Watson2015}.

Once a $g-r$ value from SDSS is assigned to each galaxy, the SED associated with that $g-r$ value is mapped onto the simulated galaxy as well. The galaxy SEDs are represented as five SED template coefficients, $\alpha_{i}$, using the templates determined used in the \textsc{kcorrect} algorithm \citep{blanton_etal:03kcorr} as the basis. Since some tolerance in $M_{r}$ is allowed in the match between simulated and data galaxies, the SED obtained from the data must be normalized such that it gives the absolute magnitude originally assigned to the simulated galaxy. This is accomplished by re-normalizing the \textsc{kcorrect} coefficients such that

\begin{align}
-2.5\log_{10}\frac{\alpha_{i}^{\prime}}{\alpha_{i}} =  M_{r,\textrm{sim}} - M_{r,\textrm{train}}
\end{align}
where $\alpha_{i}^{\prime}$ are the re-normalized coefficients, $M_{r,\textrm{sim}}$ is the $r$-band absolute magnitude assigned to the simulated galaxy and $ M_{r,\textrm{train}}$ is the absolute magnitude of the matched training set galaxy. Using these coefficients, the SEDs can be integrated over filter bandpasses in order to produce observed galaxy magnitudes.

Here we emphasize a few key points of the algorithm.  First, we note that the training set SEDs are mapped to synthetic galaxies without using redshift information. When combined with the magnitude limit of our training set, this means that the number of galaxies in the training set varies strongly as a function of $M_r$, with the highest density at $M_r \sim -20$.  Additionally, no galaxy evolution models are applied. In particular, the only two ways in which we account for redshift evolution of colors are via evolution of galaxy magnitudes {\em before} the SED is selected (which generally brightens galaxies with redshift), and via redshifting of the SED to determine the observed galaxy magnitudes. We also apply a redshift- and luminosity-dependent evolution of the red fraction of galaxies as described in Appendix E.2 of \citet{DeRose2018}, although this is not important for the SDSS comparisons presented here.
We assume both that the typical rest-frame colors of galaxies are unchanged and that the color--environment--luminosity relation remains unchanged as a function of redshift. Both of these assumptions are certainly incorrect in detail. Given these limitations, the photometry that is produced at high redshift should be treated with caution. Future development will address these issues. Nonetheless, these corrections work extremely well in reproducing galaxy properties over the redshift range of SDSS.
In a a companion paper extending this algorithm to higher redshifts in the context of DES \citep{DeRose2018}, we addressed one aspect of galaxy evolution by adjusting the relative fraction of red and blue galaxies (note that this was applied to a slightly  older version of \textsc{Addseds}).

\subsection{Summary of SED Assignment Algorithm}

We briefly summarize the most important steps in our SED assignment algorithm:

\begin{enumerate}

\item Compile a training set of spectroscopic galaxies.

\item Calculate the distance to the nearest massive halo, $R_h$, for each simulated galaxy.

\item Conditional abundance match $g-r$ colors in the training set to $\widetilde{R_h}$, a noisy version of $R_h$. 

\item Use the CAM relationship to map SEDs from the training set galaxy to the synthetic galaxy.

\item Redshift the SED and convolve with filter pass bands to determine observed magnitudes.

\end{enumerate}

Applying this algorithm results in a synthetic photometric catalog down to some limiting absolute magnitude.
Note that when attempting to model a magnitude-limited survey, such as SDSS, it is necessary to generate synthetic galaxies to some limiting absolute magnitude, $M_{r,lim}(z)$, as a function of redshift.  Because of the significant scatter in the $M_r-m_r(z)$ relation due to variation in galaxy SEDs, it is generally necessary to create significantly more galaxies than necessary, cutting the catalog to the appropriate apparent magnitude limit in a post-processing step.

\section{Observed Magnitudes and Photometric Errors}
\label{sec:observe}

\begin{deluxetable*}{ccccccc}
\tabletypesize{\footnotesize}
\tablecaption{
Survey limiting magnitudes.}
\startdata
\\
\hline
Survey &  &  & Limits & & &  \\
\hline
\hline
\smallskip
DECam & $u$& $g$ & $r$ & $i$ & $z$ & $Y$ \\
\\
DES DR2 10-$\sigma$ & & 24.07 & 23.82 & 23.11 & 22.28 & 20.79 \\
\hline
\hline
SDSS & $u$ & $g$ & $r$ & $i$ & $z$ &   \\
DR8\tablenotemark{a} & 20.4 & 21.7 & 21.2 & 20.8 & 19.3 & \\
Stripe82\tablenotemark{a} & 22.1 & 23.4 & 23.1 & 22.6 & 21.2 & \\
\hline
\hline
\smallskip
VISTA & $z$ & $Y$ & $J$ & $H$ & $K_s$  \\
VHS\tablenotemark{b} & & & 20.1 & 19.7 & 19.5 & \\
VIKING\tablenotemark{b} & 21.6 & 20.9 & 20.8 & 20.2 & 20.2 & \\
VIDEO & 25.7 & 24.6 & 24.5  & 24.0  & 23.5 &\\
\hline
\hline
\smallskip
WISE & 3.4$\mu$ & 4.6 $\mu$ \\
WISE\tablenotemark{c} & 17.1 & 15.7 & & & &  \\
\hline
\hline
LSST & $u$ & $g$ & $r$ & $i$ & $z$ & $Y$ \\
LSST-1 year \tablenotemark{e}  & 24.2 & 25.8 & 25.9 & 25.2 & 24.0 & 23.15 \\
\hline
\hline
\enddata
\tablecomments{All limiting magnitudes are $10\sigma$ AB magnitudes for galaxies unless photometric errors are not provided.}
\tablenotetext{a}{Limits appropriate for SDSS model magnitudes used for color measurements.}
\tablenotetext{b}{Limits for $2''$ aperture-corrected magnitudes.  Magnitudes
  have been converted from Vega to AB such that $z_{\mathrm{AB}} =
  z_{\mathrm{Vega}} + 0.52$; $Y_{\mathrm{AB}} = Y_{\mathrm{Vega}} + 0.62$;
  $J_{\mathrm{AB}} = J_{\mathrm{Vega}} + 0.94$; $H_{\mathrm{AB}} =
  H_{\mathrm{Vega}} + 1.38$; $K_{s,\mathrm{AB}} = K_{s,\mathrm{Vega}} + 1.8$.}
\tablenotetext{c}{Limits for \texttt{MAG\_AUTO}.  Magnitudes have been
  converted from Vega to AB such that $J_{\mathrm{AB}} = J_{\mathrm{Vega}} +
  0.91$ ; $K_{s,\mathrm{AB}}  = K_{s,\mathrm{Vega}} + 1.85$~\citep{blanton05}.}
\tablenotetext{e}{Rescaled from proposed 10-year depth for $5\sigma$ point source detections.}
\label{table:magnitudes}
\end{deluxetable*}

One of the main strengths of the \addgals\ algorithm is its ability to produce photometry in a number of different bands using empirically determined SEDs. In order for these magnitudes to be useful, it is often necessary to include photometric error estimates.
These errors cause objects above the detection threshold to scatter out of our detection limits as well as causing
many more faint objects to scatter in.  Modeling these errors appropriately can thus be important for analyses that use galaxies close to the detection limits of their respective surveys, which is the case for most weak lensing analyses.

A significant challenge is the construction an appropriate error model
that is consistent between surveys.  Existing surveys report limiting
magnitudes in several inconsistent ways.  For example, some surveys
report $5\sigma$ galaxy magnitudes, some report $10\sigma$ point source
magnitudes, and still others report limiting magnitudes by measuring
the 80\% completeness limit.

To ensure that errors are consistently defined, we have instead taken a
pragmatic approach to calculate galaxy limiting magnitudes.  For all existing
surveys, we remeasure the $10\sigma$ limiting magnitude ($m_{\mathrm{lim}}$)
given the reported galaxy photometric errors using the algorithm described in \citet{Rykoff2015}. To match the full magnitude/error distribution, we
also measure an effective exposure time ($t_{\mathrm{eff}}$) and an additional
parameter ($\Sigma_{\mathrm{int}}$, described below) to encompass variations in
survey depth, seeing, and galaxy size. 

Synthetic photometric errors are calculated using a relatively straightforward method of
calculating the Poisson noise for the flux of a simulated galaxy plus the sky
noise in a particular band.  Here, the total signal from these two sources
(galaxy and sky) are given by the relation:

\begin{equation}
  \begin{split}
    S_{\mathrm{gal}} &= 10^{-0.4(m_\mathrm{gal} - ZP)}\times t_{\mathrm{eff}} \label{eq:flux_gal}\\
    S_{\mathrm{sky}} &= f_{\mathrm{sky}} \times t_{\mathrm{eff}},
  \end{split}
\end{equation}
where $m_{\mathrm{gal}}$ is the magnitude of a galaxy and $f_{\mathrm{sky}}$
is the sky noise (in a particular band), and $t_{\mathrm{eff}}$ is the
effective exposure time.  In all cases the zero-point is set to $ZP = 22.5$, and
all fluxes in the data tables are converted to nanomaggies such that:
\begin{equation}
  m = 22.5 - 2.5\log_{10}f_{\mathrm{nmgy}}.
\end{equation}
Finally, we note that the sky noise parameter, $f_{\mathrm{sky}}$, can be estimated from the
$10\sigma$ limiting magnitude $m_{\mathrm{lim}}$ and the associated
$f_{\mathrm{lim}}$ :
\begin{equation}
  f_{\mathrm{sky}} = \frac{f_{\mathrm{lim},1}^2 \times t_{\mathrm{eff}}}{100} -
  f_{\mathrm{lim},1},
\end{equation}
where $f_{\mathrm{lim},1}$ is the 1-second flux at the limiting magnitude given
by \cref{eq:flux_gal}.

Given the galaxy flux and sky flux, in the simplest form the typical noise
associated with each galaxy will be given by a random draw from a distribution
of width $\sigma_{\mathrm{flux}} = \sqrt{S_{\mathrm{gal}} +
S_{\mathrm{sky}}}$.  However, in a simple model to account for variations in
galaxy size, survey depth, and seeing, we add in an additional log-normal
scatter parameter $\Sigma_{\mathrm{int}}$.  After taking
$\sigma_{\mathrm{int}}$ as a random draw from a distribution of width
$\Sigma_{\mathrm{int}}$, we arrive at:
\begin{equation}
  \sigma_{\mathrm{flux,tot}} = \exp(\ln\sigma_{\mathrm{flux}} + \sigma_{\mathrm{int}}).
\end{equation}
For most surveys, the typical value for $\Sigma_{\mathrm{int}}$ is $\sim0.2 -
0.3$, equivalent to a 20--30\% scatter in effective depth. For particular survey applications
such as in \citet{DeRose2018}, we use maps of the effective depth variation as a function
of sky position in order to more realistically model these variations.

After taking a random draw from a distribution of width
$\sigma_{\mathrm{flux,tot}}$ for each galaxy, the total observed
flux and error are converted to nanomaggies, such that $f_{\mathrm{nmgy}} =
S_{\mathrm{gal,obs}}/t_{\mathrm{eff}}$. Finally, magnitudes and
magnitude errors are calculated as:
\begin{equation}
  \begin{split}
    m_{\mathrm{obs}} &= 22.5 - 2.5\log_{10}(f_{\mathrm{nmgy}}) \\
    m_{\mathrm{err,obs}} &= \frac{2.5}{\ln(10)}\frac{f_{\mathrm{err,nmgy}}}{f_{\mathrm{nmgy}}}
  \end{split}
\end{equation}

We provide galaxy magnitude limits for a number of existing and planned surveys. \Cref{table:magnitudes} lists all the survey magnitudes included in our simulations, including the filters and the limiting magnitudes for each filter. The existing surveys included are SDSS DR8~\citep{dr8}, SDSS
Stripe 82 coadds~\citep{annis11}; 
WISE~\citep{jarrett11}; VHS \citep{McMahon2013}, VIDEO
\citep{jarvis_etal:13}, and VIKING \citep{sutherland:12}; and DES DR2 \citep{DES_DR2}. We also
produce magnitudes for Rubin Observatory's Legacy Survey of Space and Time, where one-year limiting magnitudes
are obtained by re-scaling the projected 10-year depth for $5\sigma$ point source detections\footnote{https://docushare.lsstcorp.org/docushare/dsweb/Get/LPM-17}.

\section{Validation Against SDSS Data}
\label{sec:coltests}
We now present tests of our magnitude and SED assignment algorithms. 
Here we focus primarily on the global distribution of galaxy colors, clustering as a function of color, and radial profiles of galaxies around clusters in the local Universe, compared to data from SDSS. \citet{DeRose2018} presents additional tests of the model, including additional comparisons between our synthetic catalogs and observations of galaxy clusters, as well as comparisons of luminous red galaxy populations, high-redshift colors, and photometric redshifts in the DES. All of the comparisons presented in this section use \addgals\ run on the \textsc{L1} lightcone, producing a quarter sky ($10,313.25$ square degree) footprint out to $z=0.32$.

\subsection{One-Point Statistics}

\begin{figure*}
\includegraphics[width=\columnwidth]{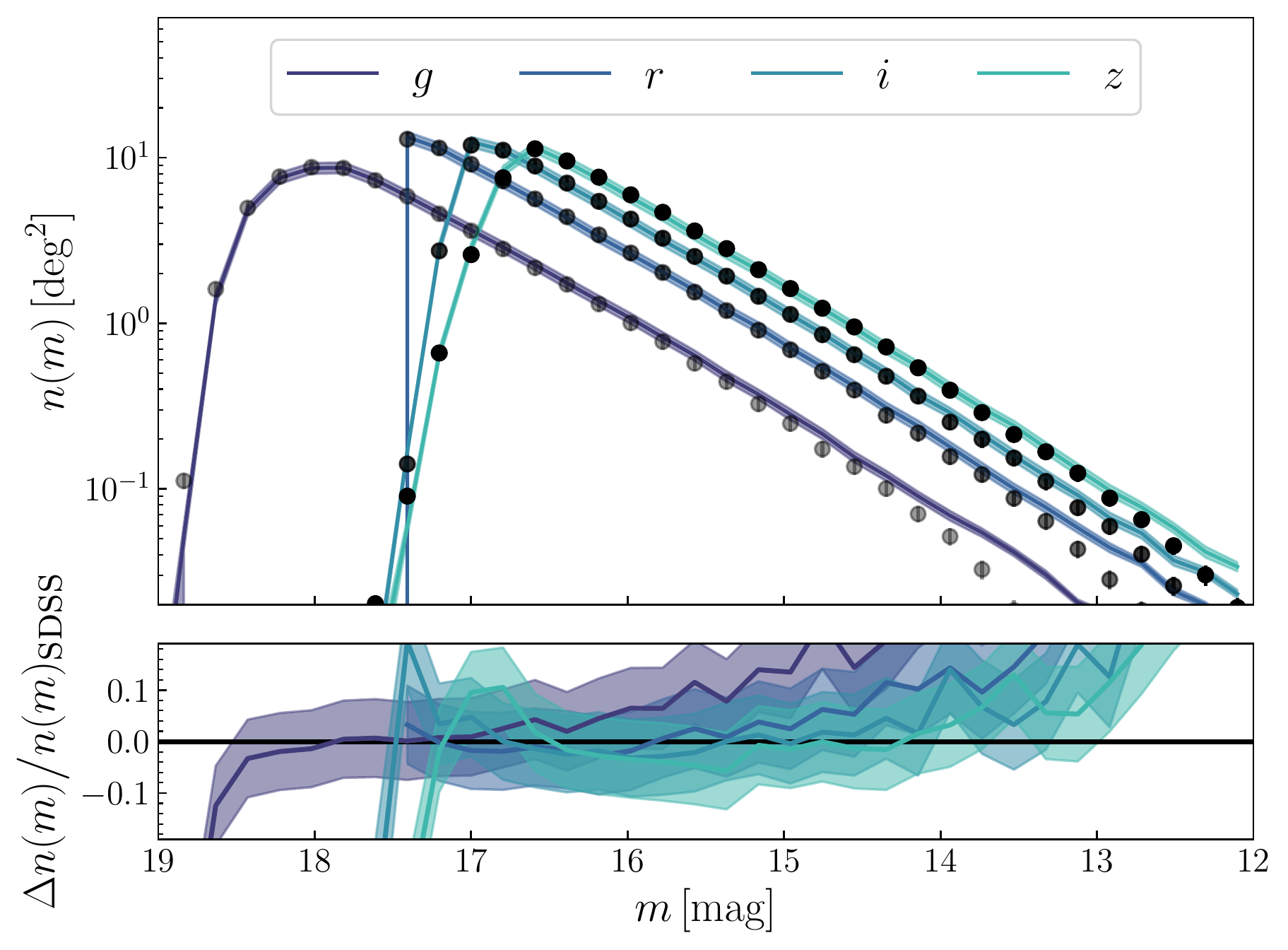}
\includegraphics[width=\columnwidth]{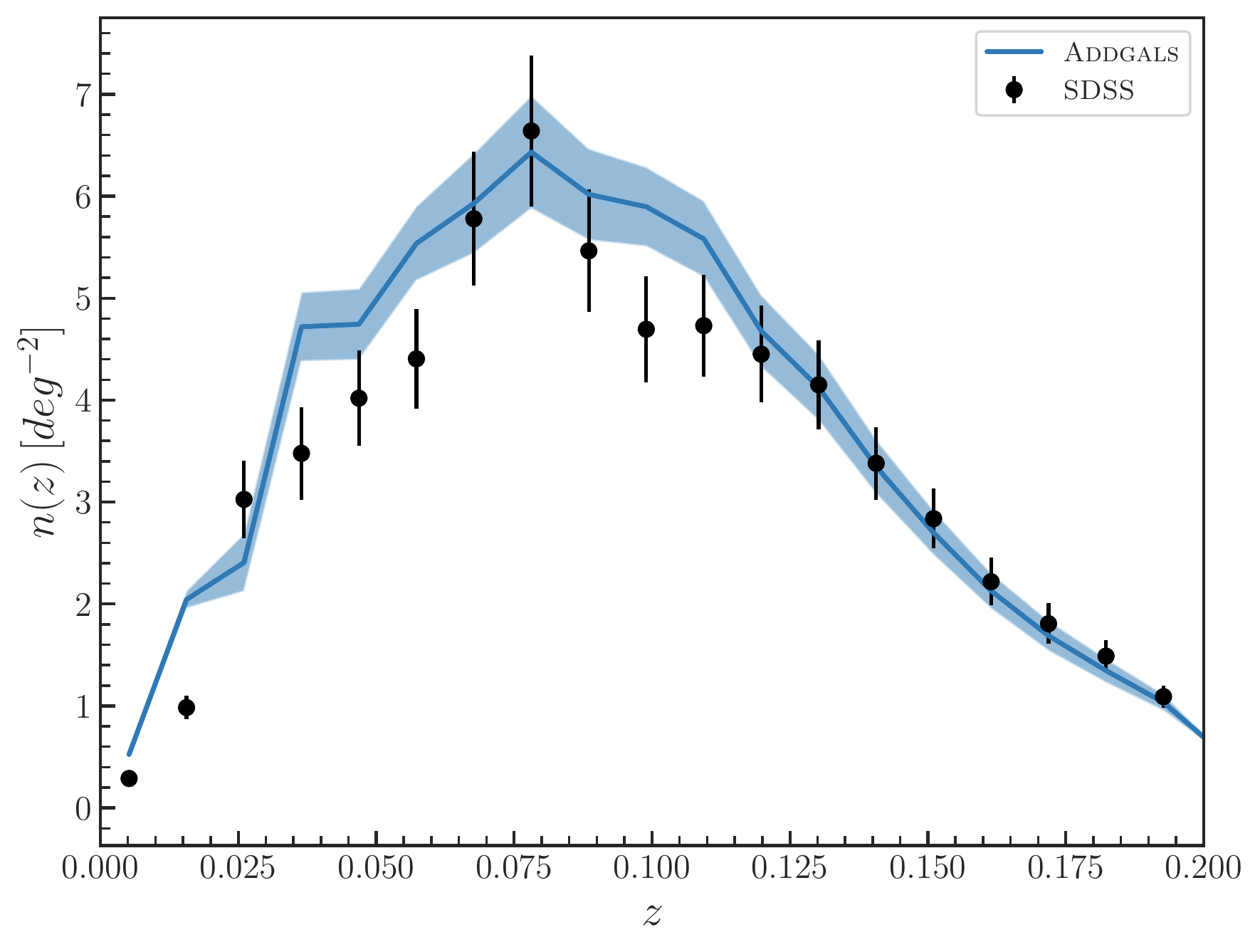}
\caption{
\emph{Left}: Galaxy counts in the SDSS $griz$ bands, for all galaxies brighter than $m_r < 17.7$ and at redshifts $z  < 0.2$.  Black points indicate the SDSS DR7 VAGC sample and lines are the L1 \addgals\ simulation. Error bars are calculated via jackknife on $\sim 200$ sq. degree regions. \emph{Right}: Redshift distribution for simulated galaxies selected to match the
SDSS spectroscopic sample, compared to the redshift distribution for galaxies with measured redshifts in the main sample of SDSS DR7. In each case, the galaxies are limited by $14.0<r<17.7$ and $0 < z < 0.2$, to match the SDSS training sample used. Error bars are calculated via jackknife on $\sim 200$ sq. degree regions.}
\label{fig:z_mag_dist}
\end{figure*}

\begin{figure*}
\includegraphics[width=\linewidth]{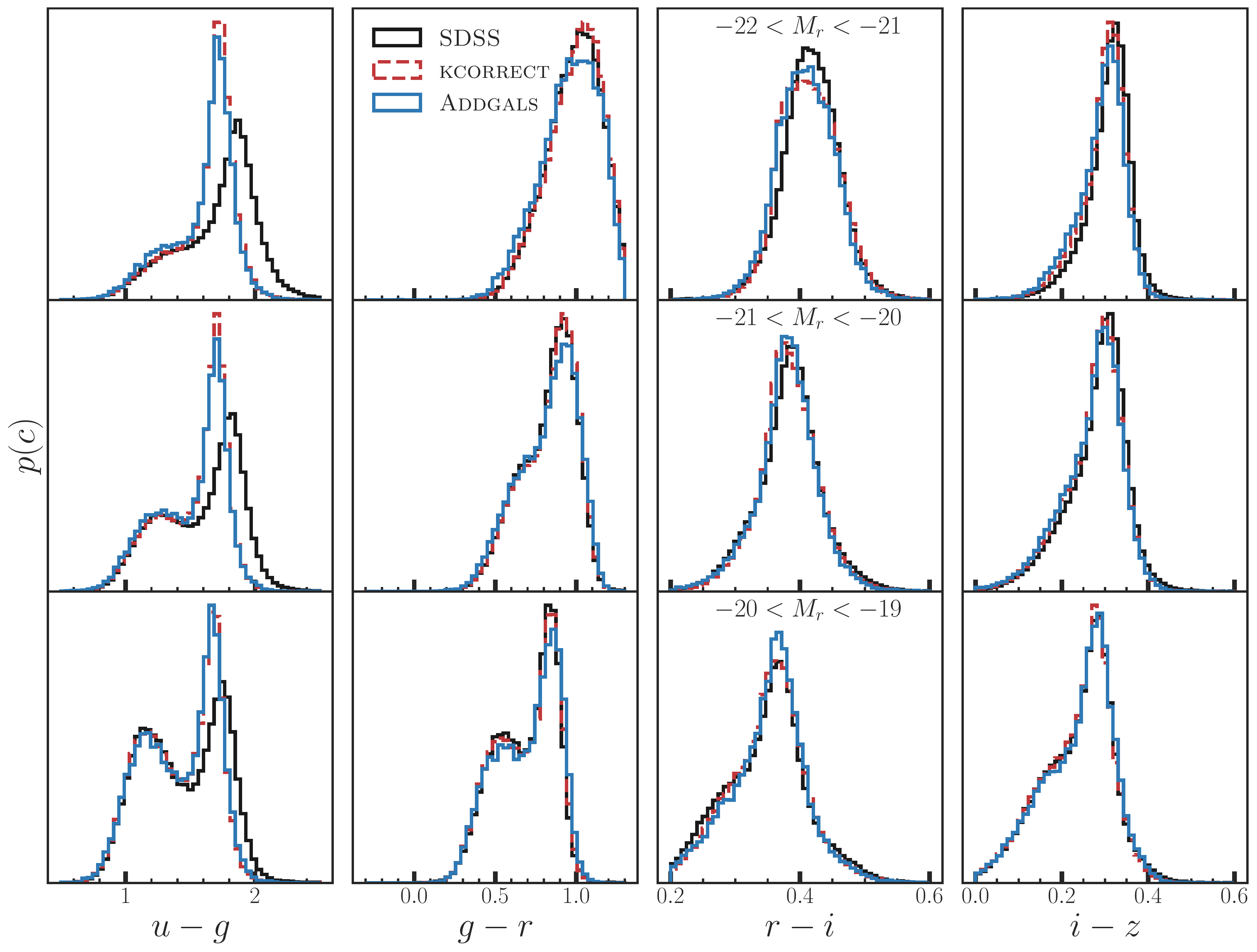}
\includegraphics[width=\linewidth]{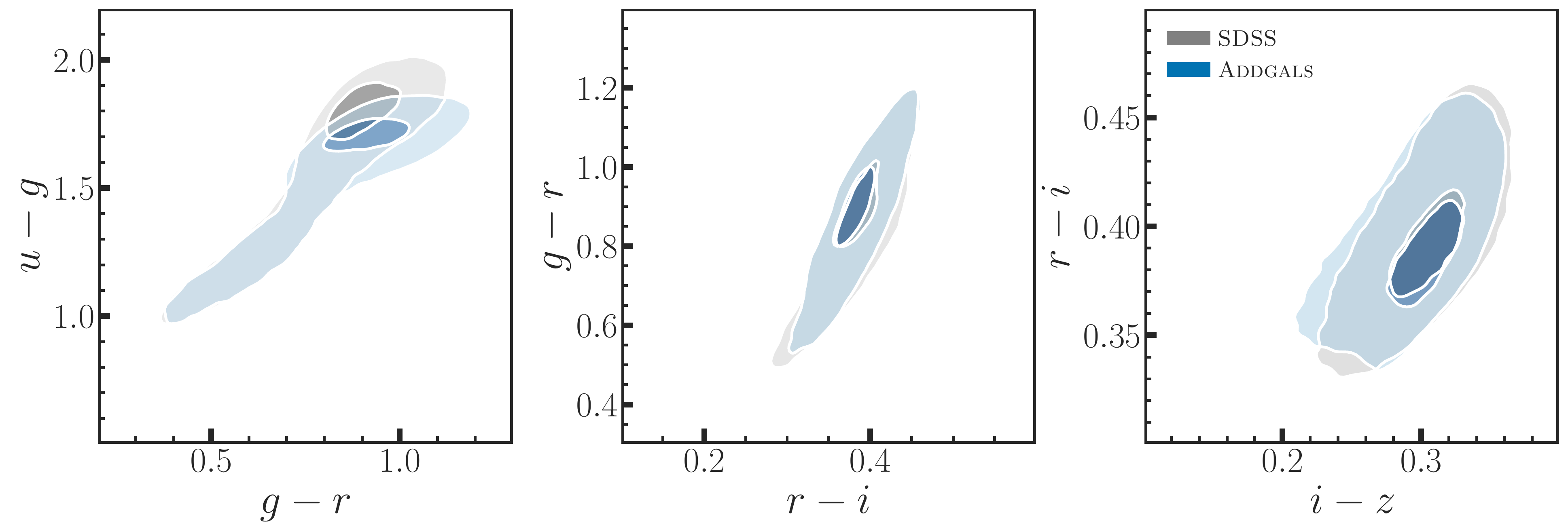}
\caption{\emph{Top}: Distributions of $u-g$, $g-r$, $r-i$, and $i-z$ colors (columns) in bins of absolute magnitude (rows). In all panels, the black line represents the distribution in SDSS DR7, while the blue lines show the distributions for \addgals\ \textsc{L1} catalog, and the red dashed lines show the colors predicted from the SDSS DR7 training set if reconstructed from their \textsc{kcorrect} fits. Nearly all discrepancies between \textsc{Addgals} and SDSS are due to inaccuracies in \textsc{kcorrect}, not our method for assigning SEDs to our simulations, as can be seen by the good match between the blue and red lines. \emph{Bottom}: Color--color distribution for galaxies with $m_{r}<17.77$. In both panels, grey contours show measurements from the SDSS DR7 main galaxy sample; blue contours show the \textsc{Addgals} catalog.  Contours include 39\% and 84\% of the galaxies.}
\label{fig:color_dist}
\end{figure*}

We first investigate the ability of \addgals\ to reproduce color, magnitude, and redshift distributions by comparing with the SDSS main galaxy sample. The left panels of \cref{fig:z_mag_dist} show histograms of observed magnitude counts in $griz$ bands. The \addgals\ catalog is compared to the magnitude-limited SDSS main sample with $m_r < 17.77$, where the error bars shown are computed with jackknife using regions of approximately $200$ sq. degrees. The agreement is good to better than $10\%$ to $r \sim 13$, with similar performance at the same number density in the other bands. The discrepancies seen on the bright end are a result of the redshift evolution imposed in our input rest-frame luminosity function in order to match DES galaxy number densities as described in \citet{DeRose2018}. The upturns at the faint end in $griz$ where SDSS is very incomplete are sensitive to the assumed photometric error model. The $u$-band (not plotted) performs significantly worse as a result of the discrepancies seen in \cref{fig:color_dist}; which we believe are largely due to the fact that the SED templates are not fully tuned in to $u$-band data.

The right panel of \cref{fig:z_mag_dist} shows the redshift distribution for galaxies in the simulated catalogs compared with those in the SDSS DR7 main sample, using the same magnitude and redshift cuts as the previous comparison. Error bars are computed using the same jackknife procedure. Again we find good agreement.

The top section of \cref{fig:color_dist} presents the distributions of observed $u-g, g-r, r-i$, and $i-z$ as a function of $M_r$ in our simulations (blue), compared with those from our training set (black). Also displayed in red are the distributions that are obtained when reconstructing each training sample galaxy's magnitudes using only their \textsc{kcorrect} coefficients. The same magnitude and redshift cuts applied to the training set are also applied to the \addgals\ catalog. Different rows in this figure show bins of absolute magnitude as indicated by the labels. 

Although the SEDs from \addgals\ are selected from a training set of these SDSS galaxies, matches in the global color distribution are not guaranteed. The reason for this is twofold. First, if the joint distribution of $M_r$ and $z$ found in SDSS is not reproduced in our simulations, then even if $p({\rm SED} | M_r)$ is perfect, the observed colors will not be matched by the simulation. Secondly, \textsc{kcorrect} coefficients are a lossy compression of galaxy SEDs, and so when these SED representations are integrated over bandpasses, they are not guaranteed to exactly reproduce observed magnitudes and colors.

For $g-r$, $r-i$ and $i-z$ we find very good agreement between SDSS and \addgals\ for all magnitude bins. For $u-g$ the agreement between \addgals\ and SDSS is significantly worse for all magnitudes, with \addgals\ showing a narrower red-sequence that is slightly shifted to low $u-g$ relative to the data. The reason for this discrepancy is that the \textsc{kcorrect} coefficients that are used to represent the training set SEDs are not fit to the $u$-band in SDSS. This means that these SED fits do a worse job at reproducing $u-g$ colors, even when comparing the colors predicted from the template fits to the observed colors on a galaxy-by-galaxy basis in our training set \citep{blanton_etal:03kcorr}. This is evidenced by the fact that the red lines also show the same disagreement with black. The bottom section of \cref{fig:color_dist} shows joint distributions of $u-g$ and $g-r$, $g-r$ and $r-i$, and $r-i$ and $i-z$ colors. Again we see worse performance in $u-g$ due to the aforementioned issues with \textsc{kcorrect} model fits, but otherwise the agreement is very good.

\subsection{Color-Dependent Clustering}
\label{sec:color-env}

We have shown previously in \cref{sec:lumtests} that the
correlation function of galaxies at a given absolute magnitude can be well
matched using \addgals. Here we test whether the SED assignment algorithm is able 
to reproduce clustering as a function of magnitude and color by splitting our
simulated samples into red and and blue sub-samples and comparing again to the SDSS DR7
main galaxy sample.

\begin{figure*}
\centering
\includegraphics[width=1.75\columnwidth]{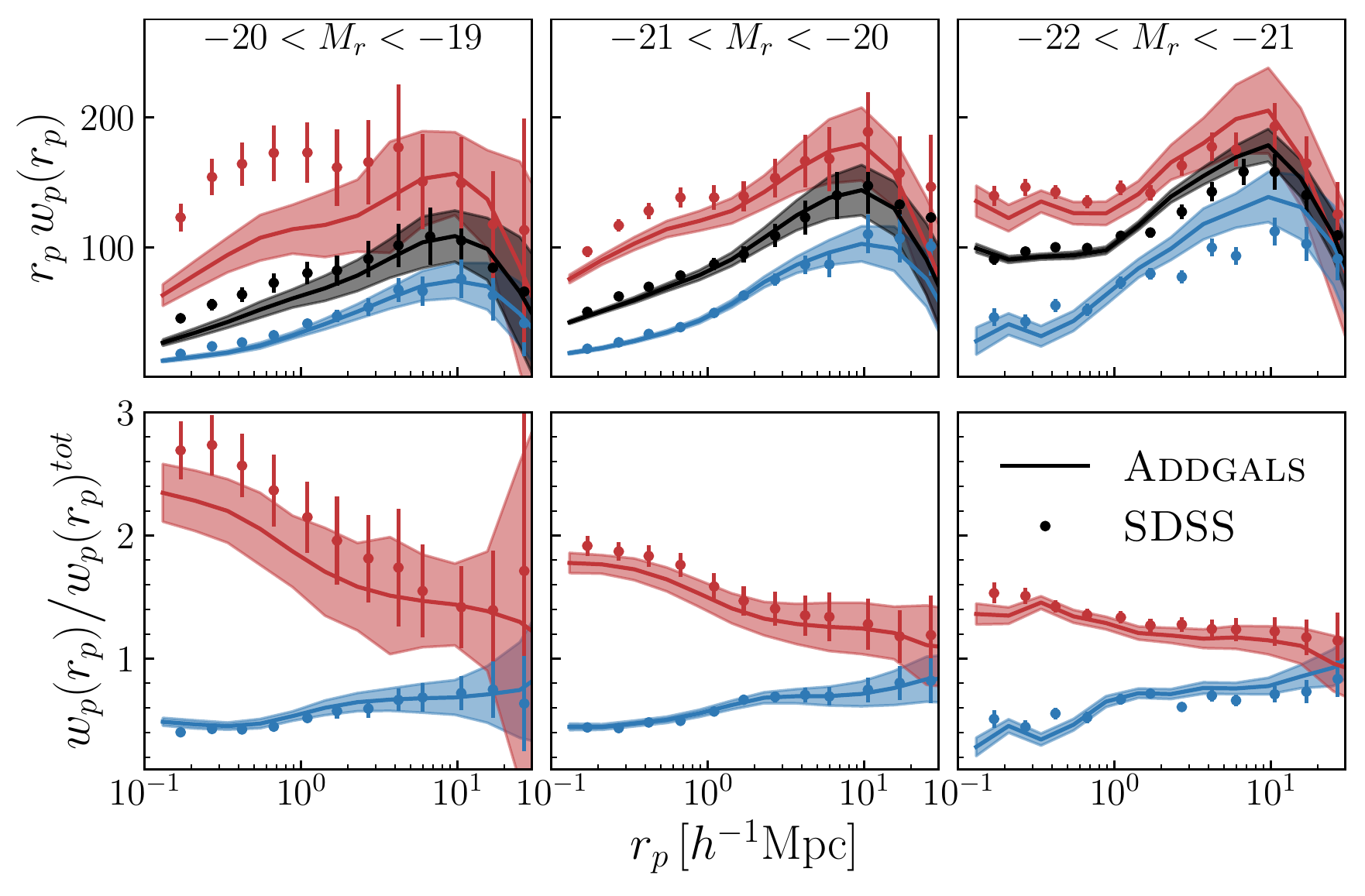}
\caption{
\emph{Top}: Projected galaxy correlation function in magnitude-selected samples for \addgals\ applied to the L1 simulation (lines) compared to the measurements from SDSS (\citealt{Zehavi11}; points). Correlation functions binned by $M_r$ only are shown in black; red and blue galaxies are shown in red and blue.  \emph{Bottom}: The red and blue clustering measurements as shown in the top panel divided by the same measurements without color selection for the \addgals\ L1 simulations and the data.}
\label{fig:color_cf} 
\end{figure*}

In \cref{fig:color_cf} we compare our simulations to projected clustering measurements in magnitude bins from \citet{Zehavi11}, adhering to their definition of red galaxies: $g-r > 0.21 - 0.03 M_r$. We employ the same $w_p(r_p)$ estimator procedure as outlined in \cref{sec:lumtests}, but now distributing the randoms uniformly over the unmasked $10,313$ square degrees covered by our lightcone simulations, drawing redshifts for our random points to match the distribution of redshifts followed by each galaxy sample separately. The same redshift binning as employed by the SDSS measurements is used for each magnitude bin. Errors on our simulations are estimated via jackknife using $\approx 200$ square degree regions.

The trends in the SDSS data are reproduced by our simulations, with red galaxies significantly more clustered than their blue counterparts at fixed $M_r$. The discrepancies between the red and blue galaxy clustering measurements in our simulations and those in SDSS are largely a consequence of issues with the clustering as a function of $M_r$, especially for the faintest sample shown. In the bottom panel of this figure, the measurements in the top panel are divided by the $w_p(r_p)$ measurements for samples with the same absolute magnitude selection, but without color selection in order to remove discrepancies caused  by issues in the model as a function of $M_r$ alone. Although the fit is not good in a chi-squared sense, we see that most of the discrepancy in the top panel is due to imperfect modeling of clustering as a function of $M_r$, not our SED assignment algorithm, although red galaxies are still slightly under-clustered with respect to the SDSS measurements.

\begin{figure}
\includegraphics[width=\columnwidth]{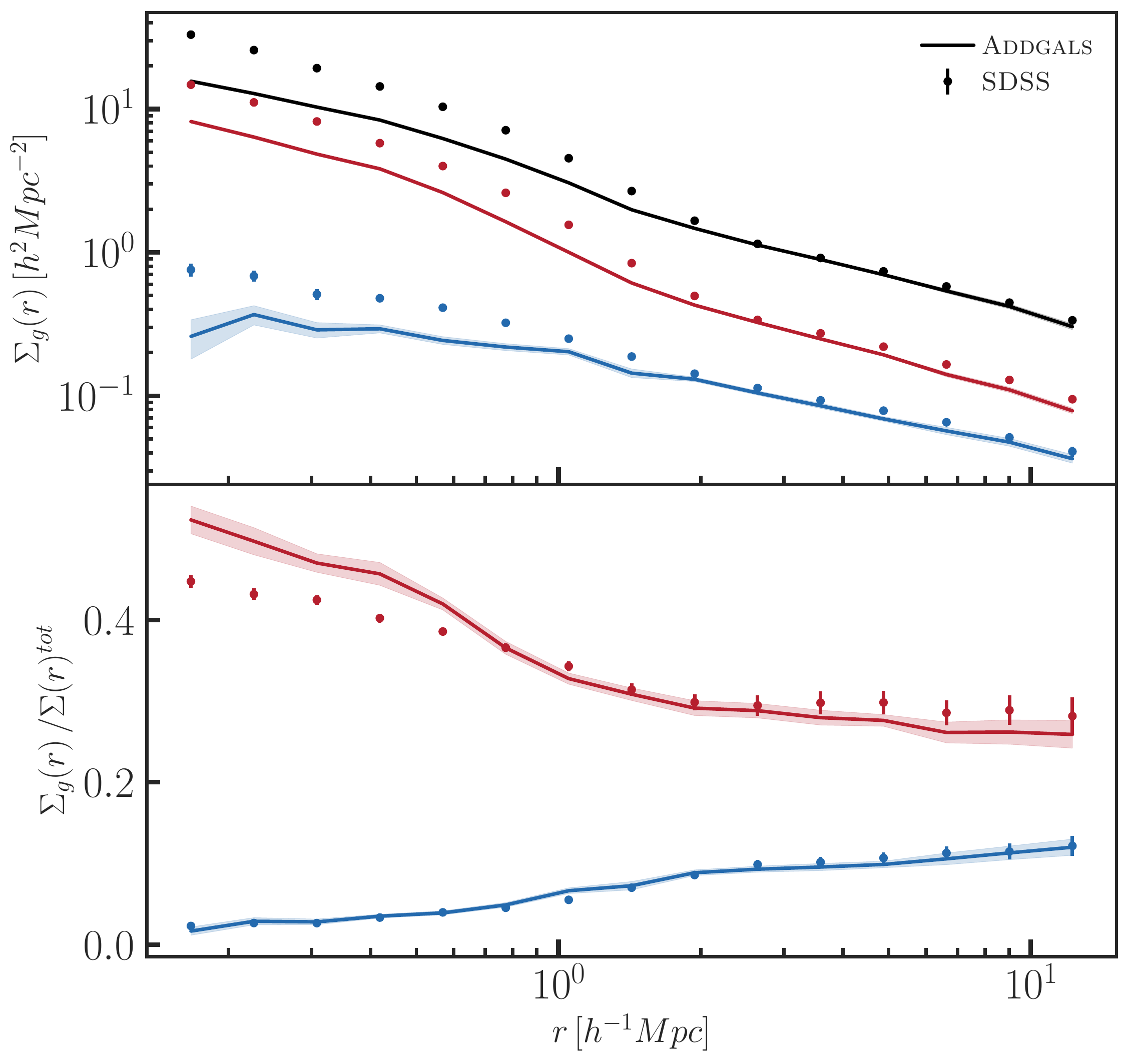}
\caption
{\emph{Top}: Projected galaxy profiles around \textsc{RedMaPPer} clusters for red, blue, and all galaxies with $M_r<19.43$. Profiles measured in L1 \addgals\ are compared to measurements from \cite{Baxter2015}. The deficit in clustering that is seen for all samples is likely due to artificial subhalo disruption in the SHAM model that the \textsc{Addgals} model is trained on. This effect is important at larger scales in this measurement than in \cref{fig:wpcomp}, because it includes
higher mass halos. \textit{Bottom}: 
Projected galaxy profiles split by color, normalized by the profile for all galaxies. 
The trends in color are well captured by \addgals\, although the quenched fraction $f_q(r)$ is slightly over-predicted at small scales.}
\label{fig:color_sigmag}
\end{figure}

Despite the reasonable performance exhibited in figure \ref{fig:color_cf}, \addgals\ is not able to reproduce the abundance of galaxy clusters as a function of richness, $\lambda$, a common mass proxy used in analyses of \textsc{RedMaPPer} clusters \citep{rozo_etal:11, rykoff_etal:13}. To see why this is, \cref{fig:color_sigmag} compares projected galaxy profiles around \textsc{RedMaPPer} galaxy clusters between \addgals\ and SDSS. The SDSS measurements are taken from \citet{Baxter2015}, and our measurements use the same procedure as detailed there. The only difference between our measurements and the measurements in \citet{Baxter2015} is that the richness cut made on the cluster catalog has been adjusted to $\lambda > 9.3$ rather than $\lambda > 20$ in order to match the abundance of clusters found in SDSS. In doing so, this figure examines galaxy profiles around halos of similar masses in the simulations and SDSS data. Profiles around clusters of the same richness show much better agreement, mostly due to the constraint that equal richness imposes on the projected galaxy number densities at the cluster boundary. 

The top panel shows galaxy profiles for three different samples, all galaxies with $M_r < -19.43$ in black, and galaxies in the top and bottom quartiles of rest frame $g-r$ in red and blue respectively. At large scales all profiles agree quite well with the measurements in SDSS, evidencing the fact that the number densities and biases of these samples in SDSS and \addgals\ are very similar. On small scales, all three samples in our simulations exhibit much shallower profiles than seen in the data. The deficits in the red and blue samples are driven entirely by the lack of galaxies in general on these scales, and not an issue with the quenched fraction of galaxies as a function of radius $f_q(r)$. This can be seen more explicitly in the bottom panel, where we divide the red and blue galaxy profiles by the total profiles for the simulations and data respectively. Here we see that the $f_q(r)$ is actually over-predicted in our simulations for the halo mass range probed by these measurements.

The reason for the deficit in the total galaxy profile is likely artificial subhalo disruption in the \textsc{T1} simulation, which is then inherited by the \addgals\ model via our training process. Higher-resolution simulations, or an orphan model in the \textsc{T1} simulation may help to remedy these issues. Indeed, \citet{DeRose2021} shows that the inclusion of a model for orphan galaxies can significantly improve the ability of SHAM to fit $M_r<-19$ clustering measurements. This improvement is facilitated by a large increase in galaxy occupation for $M_{\rm vir} > 10^{13}\, \hmsun$, and as such would also remedy the galaxy number density issues in \addgals\ at the cluster mass scale. This increase in satellite fraction boosts large scale bias, but this is compensated for by a decrease in assembly bias which has the competing effect of decreasing large-scale bias, while largely maintaining the one-halo clustering signal (see, e.g. the top left panel of figure 2 in \citealt{DeRose2021}). Work is ongoing to incorporate orphan galaxies into the SHAM models that \addgals\ is trained on, but until these improvements are fully implemented the cluster properties in \addgals\ catalogs must be treated with caution.

\section{Resolution requirements of \textsc{Addgals}}
\label{sec:res_requirements}

Now that we have presented and validated the \addgals\ algorithm, it is important to understand its resolution requirements, as the relatively modest resolution requirements of the \textsc{Addgals} algorithm are one of its major strengths. The left side of \cref{fig:wpcomp} compares the projected clustering of \addgals\ run on the $z=0$ snapshots of three simulations with progressively lower mass resolution, \textsc{T1}, \textsc{L1} and \textsc{L2}, where the measurements and errors are computed as described in \cref{sec:lumtests}. In all cases, simulations are converged with respect to the errors on the measurements in the \textsc{T1} simulation, which has a similar volume to the SDSS main galaxy sample. The \textsc{L1} and \textsc{L2} models are also in relatively good agreement, although they show discrepancies at the $10\%$ level on small scales for the $M_{r}<-22$ sample and $5\%$ for $M_{r}<-21$ and $M_{r}<-20$ on scales $r_p<1\,\hmpc$.

\begin{figure*}
\centering 
\includegraphics[width=\columnwidth]{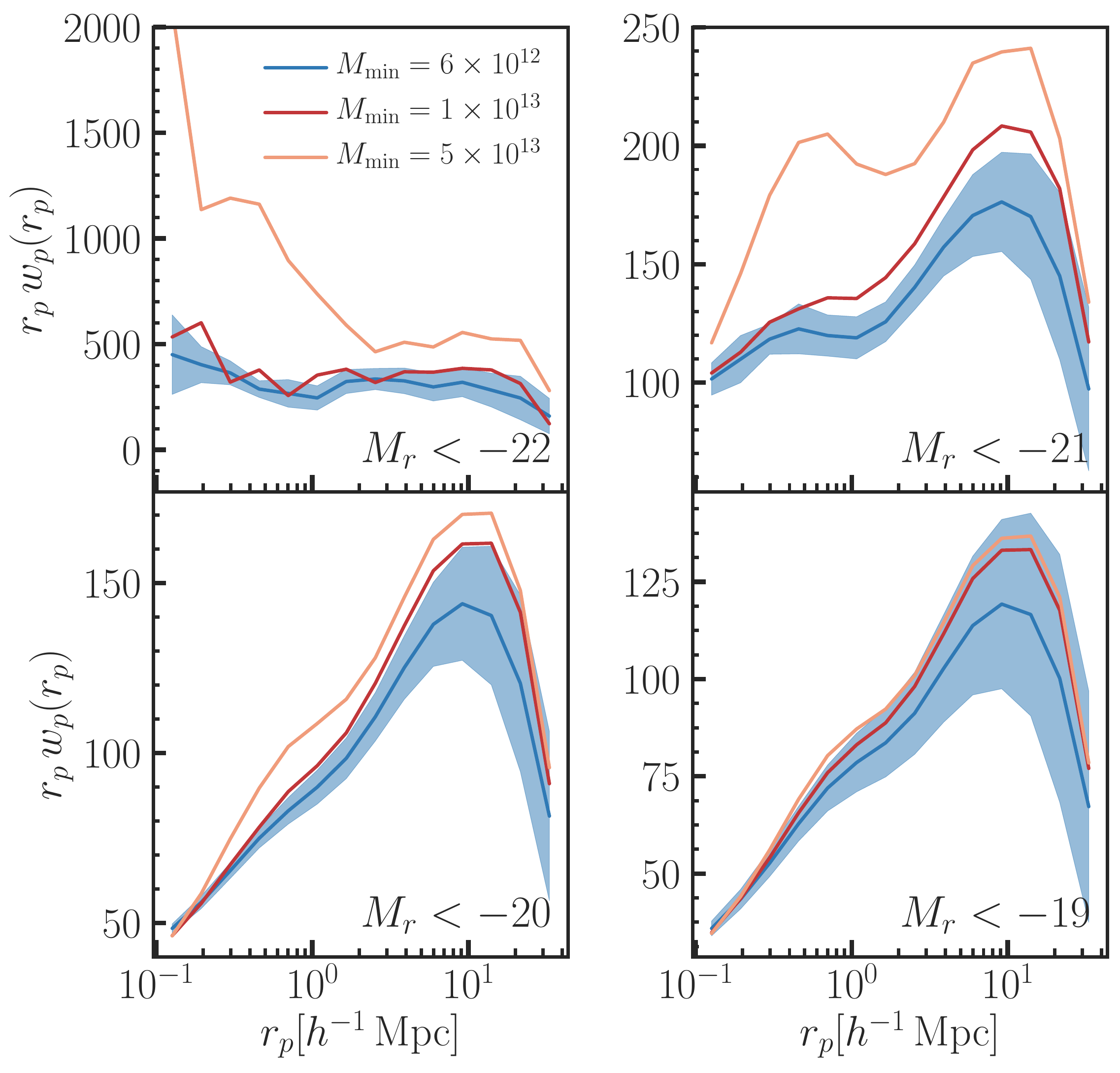}
\includegraphics[width=1.04\columnwidth]{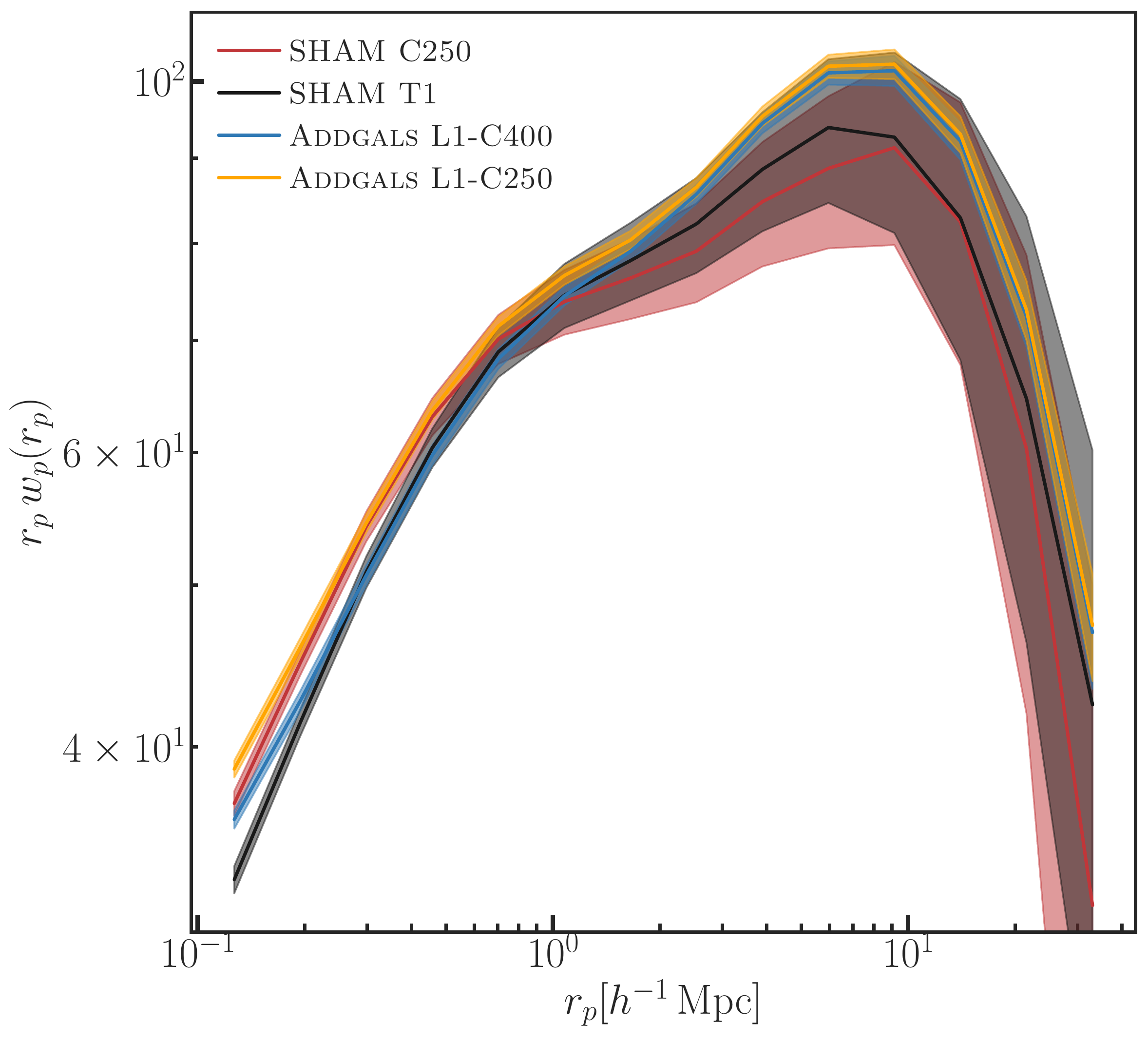}
\caption{Impact of resolution on galaxy clustering measurements in \addgals.
\emph{Left}: Projected correlation functions for absolute-magnitude-limited samples of $M_r<-22, -21, -20, -19$ for \addgals\ models, varying the minimum mass halo mass, $M_{min}$, used to populate central galaxies. This is the parameter that our models are most sensitive to, and largely drives the resolution requirements for \addgals. \emph{Right}: Projected correlation functions for $M_r<-19$ samples, varying the resolution of the input simulation used for the SHAM model. The default \textsc{T1} simulation (black) is compared to a higher resolution \textsc{C250} simulation (red); the latter model has stronger clustering below $\sim 1\hmpc$ due to reduced artificial subhalo disruption. The \addgals\ model trained on this simulation, \textsc{Addgals L1-C250} (yellow) also inherits this increased clustering compared to the default \textsc{Addgals L1-C400} model (blue).}
\label{fig:restests}
\end{figure*}

The discrepancies between \addgals\ and the \textsc{SDSS} data for the $M_r<-19$ sample are not due to an insufficiency of the \addgals\ algorithm, but rather resolution effects in the \textsc{SHAM} catalog that \addgals\ is trained on. This is demonstrated in the right side of \cref{fig:restests}, where \addgals\ models trained on two different \textsc{SHAM} models are compared. One is our fiducial \textsc{SHAM} run on the \textsc{T1} simulation. The other is a \textsc{SHAM} run on the higher resolution \textsc{C250} simulation, which was run with the same settings as the \textsc{T1} simulation, but with a simulation volume of $(250 \hmpc)^3$, $2560^3$ particles and a force softening of $\epsilon=0.8\hkpc.$ The clustering of the \textsc{SHAM C250} model is increased on small scales due to reduced subhalo disruption in the \textsc{C250} simulation relative to \textsc{T1}. We don't compare to the \textsc{SDSS} data here because the \textsc{C250} simulation is too small to use the same line-of-sight projection length of $\pi_{max}=60$ for $w_{p}(r_{p})$ as used in the data. Instead we use $\pi_{max}=20$ for this comparison. Nonetheless it is apparent that the \textsc{SHAM C250} model would agree with the \textsc{SDSS} measurements on small scales. The \addgals\ model trained on the \textsc{SHAM C250} also inherits this increased clustering on small scales. This suggests that with a sufficiently high resolution training simulation, or an orphan model that traces substructure effectively until it is physically disrupted, \addgals\ could reproduce the small-scale clustering of a $M_r<-19$ sample using a simulation with the resolution of the \textsc{L1} or even \textsc{L2} simulation.

In practice, the more important parameter governing the convergence of the \addgals\ method for projected two-point functions is the minimum mass to which central galaxies are populated, $M_{min}$. This can can be seen in on the left side of \cref{fig:restests}. All catalogs included in this figure are run on the \textsc{T1} simulation, varying $M_{min}$ between $6\times 10^{12}$ and $5\times 10^{13}\hmsun$ as indicated in the legend of the figure but keeping all other parameters fixed. In most cases the shifts are smaller than the errors on the measurements, but it is clear that this parameter is a very important free parameter of the \addgals\ algorithm, especially for brighter samples. The dependence on this parameter hints at a breakdown in the bright galaxy regime of our assumption that matching $p(R_{\delta} | M_r,z)$ of a \textsc{SHAM} is sufficient to reproduce $w_p(r_p)$ in that \textsc{SHAM}, since even for a high-resolution simulation such as \textsc{T1}, significant discrepancies in clustering are produced when too many bright galaxies are populated using this relation, whereas faint galaxies are much less affected. This breakdown is sourced by the fact that $p(R_{\delta} | M_{\rm vir}, M_r, \rm central)$ does not evolve significantly for $M_{\rm vir}>10^{13}\,\hmsun$ for bright galaxies. As such, if we do not place centrals in halos by $M_{\rm vir}>10^{13}\,\hmsun$, then using $p(R_{\delta} | M_r, z)$ induces significant scatter in $M_{vir}$, placing galaxies that should have been centrals of lower mass halos as satellites in high mass halos, thus significantly boosting the clustering signals for bright galaxies as seen in the left hand side of \cref{fig:restests}. The exact halo mass that we must populate central galaxies down to very likely depends on the mass used to compute $R_{\delta}$, but we have not explored this dependence in detail.

Finally, we expect \addgals\ to fail for simulations where the mass scale used to define \rd\ is too coarsely resolved by the particle resolution. In this regime, the density estimates produced by \rd\ will not consistently measure similar environments in the training volumes versus the volumes used for the synthetic catalogs. As simulations of this low resolution are increasingly irrelevant due to advances in computational power, we have not explored this effect in detail.

\section{Conclusions}\label{sec:conclusions}

We present the \addgals\ (Adding Density-Determined Galaxies to Lightcone Simulations) algorithm, designed to 
produce realistic simulated galaxy populations with only a modest computational cost. To achieve this goal, we employ a combination of empirical models of galaxy--halo connection in high-resolution simulations with a custom, physically motivated machine learning model that is trained to place galaxies into lower resolution volumes. This combination of techniques, which explicitly incorporates key statistical information from the data (e.g., the luminosity function and the distribution of colors/SED types), lends a baseline level of realism to the output catalog. In this work, we show that we are able to match several characteristics of the input training catalog, including matching the clustering properties of the input empirical model as a function of $r$-band absolute magnitude and redshift. We also demonstrate that we can produce realistic color distributions and can reproduce the most significant trends in clustering as a function of color. Several further comparisons are presented in \citet{DeRose2018}, which additionally describes associated weak lensing catalogs, additional redshift evolution, and tests photometric redshift and cluster finding methodology in higher-redshift synthetic surveys.

The modest simulation requirements of this method have enabled us to produce volumes of synthetic sky surveys that would be significantly more computationally expensive with other methods in active use. This includes, for example, the ability to produce the large and deep sky areas appropriate for modeling modern photometric surveys like DES, LSST, Euclid, and the Roman Space Telescope surveys. The catalogs created with the \addgals\ method have been used for a wide variety of applications including tests of photometric redshift, clustering, weak lensing, cross-correlation, and cluster finding methodology. In companion papers, the modest computational cost has allowed us to produce a significant number of such catalogs, e.g. tens of full area and depth realizations, which have be used to statistically test the performance of precision cosmological probes in the DES \citep{MacCrann17,DeRose2018, DeRose2021}.

One of the distinguishing features of \addgals' machine-learning model is that it uses a hand-crafted 
parameterization as opposed to a more generic functional form (e.g., a tree ensemble, neural network, etc.). 
This parameterization is constructed specifically to fit the simulation data, especially near the tails of 
the distribution where some extrapolation is needed. Looking ahead, generalizations of the \addgals\ 
algorithm that employ more traditional machine-learning models may be able to achieve better performance, 
but it is likely special attention will need to be paid to how these models extrapolate beyond the training 
data. This lesson is likely quite general for any machine learning model of this type, given that high-resolution simulations typically sample less volume of the universe than lower-resolution ones.

This method is not without limitations. In particular, it has difficulty precisely reproducing the small-scale clustering measured from \textsc{SDSS} of the faintest galaxy samples considered in this paper. This issue is 
likely inherited from artificial subhalo disruption in the simulation that \addgals\ is trained on. This deficit in clustering leads a low normalization of the HOD, and a deficit of galaxies at the cores of cluster mass halos with respect to observations \citep{DeRose2018}. The realistic properties of galaxy cluster populations do enable us to run modern cluster finders on the simulated data, but the lack of galaxies in the central regions of massive groups and clusters leads to an offset in the mass--richness relation that can hinder some use cases related to important cluster selection systematics.

Additionally, our SED assignment algorithm requires a representative sample of observed galaxies from which to draw SEDs in order to accurately reproduce galaxy colors. If there are SEDs that appear at high redshift, or fainter absolute magnitudes, that are not present in our low redshift \textsc{SDSS} training set, then the assumptions used by \addgals\ to populate SEDs will be broken. We showed here that in the local Universe, these assumptions can also be broken outside the wavelength range where the SED templates are well tuned, and we urge some caution for this reason in using bands outside of rest-frame $g$ through $Y$ bands. For a discussion of the extent to which these assumptions are broken in DES, see \citet{DeRose2018}. We expect that significant progress can be made in these areas especially using data from larger, deeper spectroscopic surveys to train the model.

A final significant issue with this methodology compared to more accurate methods based on fully resolved halo substructures and their histories (including hydrodynamical models, semi-analytic models, or empirical models 
based on high-resolution merger trees) is that it may lack important correlations that are expected in such models. These could include, for example, the correlated properties of both central and satellite galaxies with each other and with the larger-scale environment.

Given the size of ongoing and upcoming surveys, and their demands for accurate reproduction of galaxy magnitudes, colors, and spatial clustering, it is likely that techniques which combine empirical methods with machine learning methods in order to reduce computational cost will remain a necessary tool for precision cosmology for the foreseeable future. They will also provide a very useful complement to higher-fidelity simulations that can be produced over smaller volumes. It is thus worth considering how to mitigate some of the limitations discussed above, and active work in each of these areas is ongoing.

\addgals\ data, including a one-quarter sky simulation out to $z=2.35$ to a depth of $r=27$, with magnitudes appropriate for modeling several surveys including SDSS, DES, VISTA, WISE, and LSST, will be available upon publication at \url{ http://www.slac.stanford.edu/~risa/addgals}. 

\acknowledgements
RHW thanks her many collaborators for near infinite patience on the completion of this paper, which was begun in another era. We thank Rachel Reddick, Alex Ji, and our collaborators on the maxBCG team and in the DES collaboration, especially Chihway Chang, Carlos Cunha, Joerg Dietrich, Sarah Hansen, Brandon Erickson, Daniel Gruen, Benjamin Koester, Niall MacCrann, Chris Miller, Eduardo Rozo, Erin Sheldon, Tim McKay, and Molly Swanson, for significant useful feedback on several earlier versions of these catalogs. We thank Andreas Berlind, Derek Bingham, Joanna Dunkley, Andrew Hearin, Andrey Kravtsov, Yao-Yuan Mao, Hiranya Peiris, Eduardo Rozo, Frank van den Bosch, and Martin White for useful discussions about methodology and statistical inferrence during early development. This work received support from the U.S. Department of Energy under contract number DE-AC02-76SF00515 at SLAC National Accelerator Laboratory, and a Terman Fellowship at Stanford University. JD is supported by the Chamberlain Fellowship at Lawrence Berkeley National Laboratory. Argonne National Laboratory's work was supported by the U.S. Department of Energy, Office of Science, Office of Nuclear Physics, under contract DE-AC02-06CH11357.

This research used resources of the National Energy Research Scientific Computing Center (NERSC), a U.S. Department of Energy Office of Science User Facility located at Lawrence Berkeley National Laboratory, operated under Contract No. DE-AC02-05CH11231. Some of the computing for this project was performed on the Sherlock cluster, and on computing resources at SLAC National Accelerator Laboratory. We would like to thank Stanford University and the Stanford Research Computing Center for providing computational resources and support that contributed
to these research results. We are grateful to Stuart Marshall and the rest of the SLAC computing team for extensive support of this work.

This study made use of the SDSS DR7 Archive (as well as earlier
versions while the model was in development), for which funding has
been provided by the Alfred P. Sloan Foundation, the Participating
Institutions, the National Aeronautics and Space Administration, the
National Science Foundation, the U.S. Department of Energy, the
Japanese Monbukagakusho, and the Max Planck Society. The SDSS Web site
is http://www.sdss.org/.  The SDSS is managed by the Astrophysical
Research Consortium (ARC) for the Participating Institutions: the
University of Chicago, Fermilab, the Institute for Advanced Study, the
Japan Participation Group, the Johns Hopkins University, Los Alamos
National Laboratory, the Max-Planck-Institute for Astronomy (MPIA),
the Max-Planck-Institute for Astrophysics (MPA), New Mexico State
University, University of Pittsburgh, Princeton University, the United
States Naval Observatory, and the University of Washington.  The
authors acknowledge the support and stimulating environments of the
Aspen Center for Physics and the Kavli Institute for Theoretical
Physics (under NSF Grant No. PHY99-07949) where some of this work was
performed.

\begin{appendices}
\crefalias{section}{appendix}

\section{Constructing the Training Galaxy Catalog with Subhalo Abundance Matching}
\label{sec:shamcat}

In this work, we use subhalo abundance matching \citep[SHAM, e.g.][]{conroy_etal:06, Behroozi10, wetzel_etal:10, Reddick12, Lehmann2017} to construct the training data for the \addgals\ model. Specifically, we employ 
the model described in \cite{Lehmann2017}, placing galaxies into resolved halos 
and subhalos by matching the number density of galaxies as a function of absolute magnitude 
with that of the dark matter halos or subhalos as a function of 
$v_{\alpha} = v_{vir} \left ( \frac{v_{max}}{v_{vir}} \right ) ^{\alpha}$. The quantities $v_{max}$ and 
$v_{vir}$ are evaluated in this equation at the time when the halo is accreted onto a larger halo. 
We take $\alpha=0.684$ and $\sigma( M_{r} | v_{\alpha})=0.425$. The parameter $\alpha$ can be 
thought of as determining the concentration dependence of a halo's rank ordering, 
with larger $\alpha$ giving higher concentration halos a higher rank at fixed mass. 
The choice to evaluate the velocities used to calculate $v_{\alpha}$ at the epoch 
when the halo is accreted onto a larger halo is based on the idea that a galaxy's 
stellar mass should be much less susceptible to stripping than the outer regions 
of its dark matter halo \citep{conroy_etal:06,Reddick12}.

SHAM models generally require a single observational input, the redshift-dependent 
galaxy luminosity function (LF) in a given band. We find that a pure
Schechter function is insufficient to model galaxy luminosities for
our purposes.  At bright luminosities, there are significantly more
galaxies than a pure exponential model would predict \citep[see, i.e.,][]{Blanton03,Bernardi2013}.
In particular, the steep bright-end slope of a Schechter function results
in a very flat mass--luminosity relation for brightest-cluster galaxies
(BCGs) when using abundance matching, a relation that is inconsistent
with observations \citep[e.g.,][]{Hansen08, Kravtsov2018, To2020}. Using a luminosity function that more closely matches observations relieves this tension.

We measure the luminosity function directly using data in the SDSS DR7 VAGC,
using the same method outlined in \cite{Reddick12}. To this measurement,
we fit a modified a double-Schechter function with a Gaussian at
the bright end, as given by
\bee
\Phi(M) & = & 0.4 \ln(10) e^{-10^{-0.4(M-M_*)}}
\phi_1 10^{-0.4(M-M_*)(\alpha_1 + 1)} + \nonumber \\
& & 0.4 \ln(10) e^{-10^{-0.4(M-M_*)}}  \phi_2 10^{-0.4(M-M_*)(\alpha_2 + 1)}) + \nonumber \\
& & {\phi_3 \over \sqrt{2\pi \sigma_{hi}^2}} e^{{-(M-M_{hi})^2 \over 2 \sigma_{hi}^2}}.
\label{eq:dsg}
\eee
At $z = 0.05$, we find that equation \ref{eq:dsg} with parameters
listed in table \ref{table:dsg} reproduces the observations extremely well.
We also include evolution in this luminosity function with redshift by allowing 
for evolution in $\phi_{i}$, $M_{*}$ and $M_{hi}$ of the form:
\begin{align}
M_{*/hi}(z) = M_{*/hi,0} + Q\left (\frac{1}{1+z} - \frac{1}{1.1}\right)\, ,
\end{align}
and
\begin{align}
\phi_{i}(z) = \phi_{i,0} + Pz.
\end{align}
The value of $P$ is taken from \citet{Cool2012}, but $Q$ is fit to match counts 
as a function of magnitude from DES Y1 data. This evolution is constrained to 
be very small over the redshift range relevant to the current work. We refer 
readers to \citet{DeRose2018} for more details related to how this evolution is constrained. 

\begin{deluxetable}{cc}
\tabletypesize{\footnotesize}
\tablecaption{Parameters of the SDSS DR7 $r-$band $z = 0.05$ luminosity function
as defined by equation \ref{eq:dsg}. \label{table:dsg}}
\startdata
\\
$\phi_1$ & $0.0156 \pm 0.03 \hinv$Mpc\\
$\phi_2$ & $0.00671 \pm 0.00029 \hinv$Mpc \\
$\alpha_1$ & $-0.166 \pm 0.041$ \\
$\alpha_2$ & $-1.523 \pm 0.01$ \\
$M_*$ & $-19.88 \pm 0.03- 5\log(h)$ \\
$\phi_3$ & $(3.08 \pm 3.24)\times 10^{-5} \hinv$Mpc\\
$M_{hi}$ & $-21.72 \pm 0.52 - 5\log(h)$ \\
$\sigma_{hi}$ & $0.484 \pm 0.192$ \\
\enddata
\end{deluxetable}

With the luminosity function above, we use SHAM to populate all 100 snapshots 
of the \textsc{T1} simulation. Our catalogs are complete down to roughly 
$M_r - 5\log(h) = -19$ in the \textsc{T1} simulation and provide an excellent 
fit to the observed SDSS magnitude dependent two-point correlation function as 
measured in \citet{Reddick12}. A comparison of the SHAM algorithm applied to 
the \textsc{T1} simulation with SDSS data is shown in Fig. \ref{fig:wpcomp},
and is described further in Sec. \ref{sec:lumtests}.

\section{Modeling $p(R_{\delta}|M_{r},z)$}
\label{app:density-modeling}

In order to determine $\Theta(x,z)$ (see \cref{eq:pdf}), $\rd$ is measured at the 
position of each galaxy in the SHAM catalogs. The function $\hat{p}(\rd |M_{r}<x_i, z_j)$ is then determined, 
where the hat denotes that this is a measured quantity in the $i$-th magnitude 
bin and $j$-th snapshot, using magnitude bins with width $\Delta M_r = 0.1$ 
between $-23 < x_i < -18$ in the 56 snapshots with $z_j < 2.5$. 
$\hat{p}(\rd |M_{r}<x_i, z_j)$ is used rather than $\hat{p}(\rd |M_r=x_i, z_j)$, because 
the former quantity is significantly less noisy for bright magnitudes and 
we have found that this allows for more robust estimation of the parameters in Eq.~\ref{eq:pdf}.

\Cref{eq:pdf} is then fit to each magnitude cut, $x_i$, and redshift, $z_j$. 
In practice, we do not fit for $\Theta$, instead opting to perform the fit in a 
basis where the parameters have minimal covariance. To achieve this, $\hat{p}(\rd |M_{r}<x_i, z_j)$ 
is first fit using the original set of parameters, $\Theta$, maximizing 
the likelihood given by
\begin{align}
\mathcal{L} = \mathcal{N}(\hat{p}(\rd |M_{r}<x_i, z_j) - p(\rd;\Theta(x,z)), \hat{\Sigma}_{i,j})\, ,
\end{align}
where $\hat{\Sigma}_{i,j}$ is the covariance matrix of $\hat{p}(\rd |M_{r}<x_i, z_j)$ between each $\rd$ bin. The covariance matrix is estimated via jackknife using 125 equal volume
sub-regions of the \textsc{T1} simulation. This procedure allows the estimation of the 
parameter covariance matrix, $\bar{T}$, as the mean of the parameter 
covariance matrices for each redshift and magnitude bin, $T_{i,j}$.

\begin{figure*}
\centering 
\includegraphics[width=0.5\linewidth]{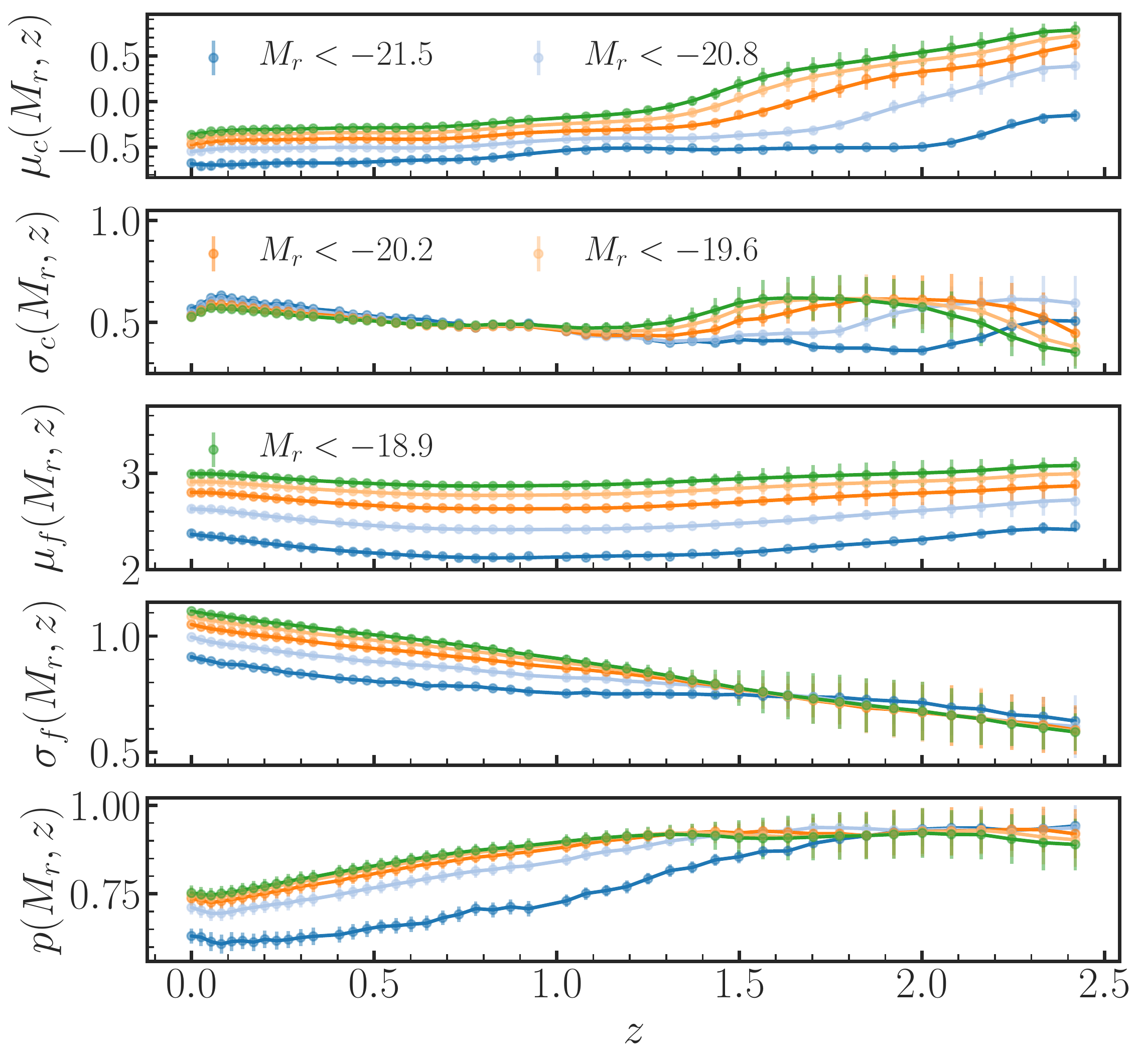}
\caption{Parameters of the model for $p(\rd|M_r,z)$, $\sigma_{c}$, $\sigma_{f}$, $\mu_{c}$, $\mu_{f}$, and $p$,  as a function of redshift and magnitude. Lines show the
Gaussian process model for this redshift and magnitude dependence.
\label{fig:gaussianprocess}}
\end{figure*}

The function $\hat{p}(\rd |M_{r}<x_i, z_j)$ is then fit to each magnitude and 
redshift bin again, this time performing the maximization over the parameter space defined by
\begin{align}
\Theta^{'} = \bar{P}\Theta\, ,
\end{align}
where $\bar{P}$ is the change of basis matrix that diagonalizes $\bar{T}$, 
yielding a set of estimated parameters for each magnitude and redshift bin, $\Theta^{'}_{i,j}$. 
In order to smoothly interpolate between these as a function of $\mr$ and $z$, a Gaussian process is fit 
to the set of $\Theta^{'}_{i,j}$. With the Gaussian process model, $\hat{\Theta}^{'}(x,z)$, it is possible to 
predict $p(\rd | M_r<x, z) = p(\rd ; T^{-1}\hat{\Theta}^{'}(x,z))$. \Cref{fig:gaussianprocess} shows the Gaussian process fits to the parameters of this model, and shows the parameter trends with redshift and magnitude.

\section{Sampling from $p(R_{\delta}|M_{r},z)$}
\label{app:density-sampling}
Here we describe how we draw samples of densities, $R_{\delta}$, from $p(R_{\delta}|M_{r},z)$, where $M_{r}$ and $z$ are the absolute magnitude and redshift of a galaxy in our simulation. It is trivial to convert random samples from a uniform distribution into samples from an arbitrary one-dimensional probability distribution function, using the cumulative distribution function (CDF) of the PDF, which can be obtained by numerically integrating the CDF. The difficulty in our case is that we don't have direct access to $p(R_{\delta}|M_{r},z)$, but rather to $p(R_{\delta}|M<M_{r},z)$, since this quantity can be measured with significantly less noise in our simulations than $p(R_{\delta}|M_{r},z)$. This is particularly true for the brightest galaxies since there are few of these in the training simulation. Because we know the average luminosity function in our training simulation, we can convert $p(R_{\delta}|M<M_{r},z)$ to $p(R_{\delta}|M_{r},z)$ using:

\begin{align}
\label{eq:density_transform}
p(R_{\delta}|M_{r}) &= \frac{1}{Z}[N(M_{r} + \delta M_{r}) p(R_{\delta}|M_{r} + \delta M_{r}) \\
& \;\;\;\; \;\;\;\;- N(M_{r}) p(R_{\delta}|M_{r})] \nonumber \\
& = \frac{\mathcal{I}(R_{\delta}|M_{r})}{Z}\, ,
\end{align}
\noindent 
where 
\begin{align}
    Z = \int d R_{\delta} \mathcal{I}(R_{\delta}|M_{r})\, .
\end{align}
We have dropped the $z$ argument to all functions for legibility, and where $N(M_r) = \int_{-\infty}^{M_r} d M_r^{\prime} \phi(M_r^{\prime})$ is the cumulative number density of all galaxies brighter than $M_r$. In practice, we evaluate $p(R_{\delta}|M_{r}, z)$ on a grid of redshift and absolute magnitude using Eq. \ref{eq:density_transform}. For the $i$th galaxy, we choose the grid point nearest to $M_{r,i}$ and $z_i$ and sample from the appropriate $p(R_{\delta}|M_{r}, z)$ to draw a density.

\section{Halo Occupation Statistics in  \textsc{ADDGALS}}\label{sec:hodstats}
\label{app:more_validation}
\begin{figure*}
\includegraphics[width=0.5\linewidth]{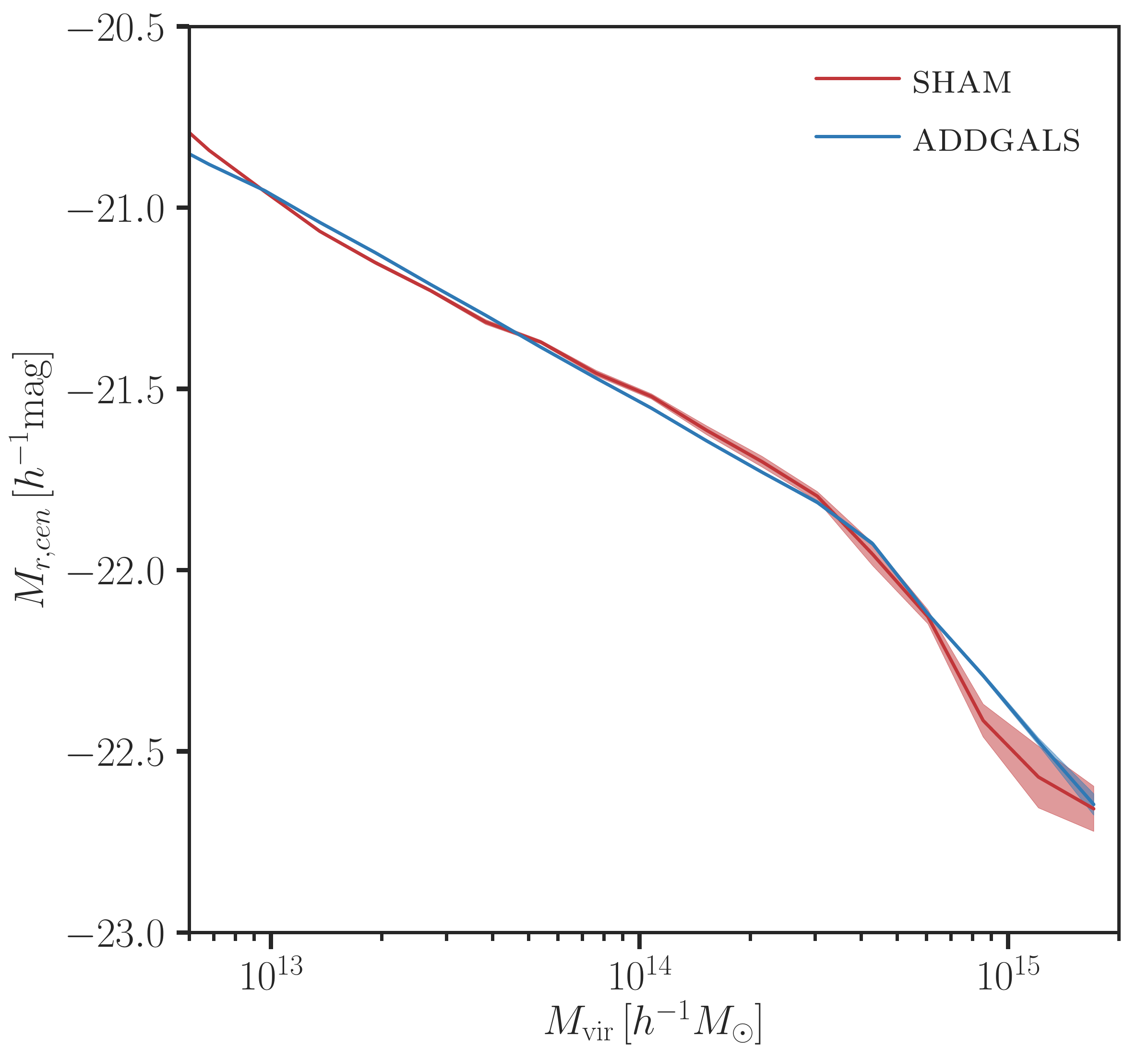}
\includegraphics[width=0.5\linewidth]{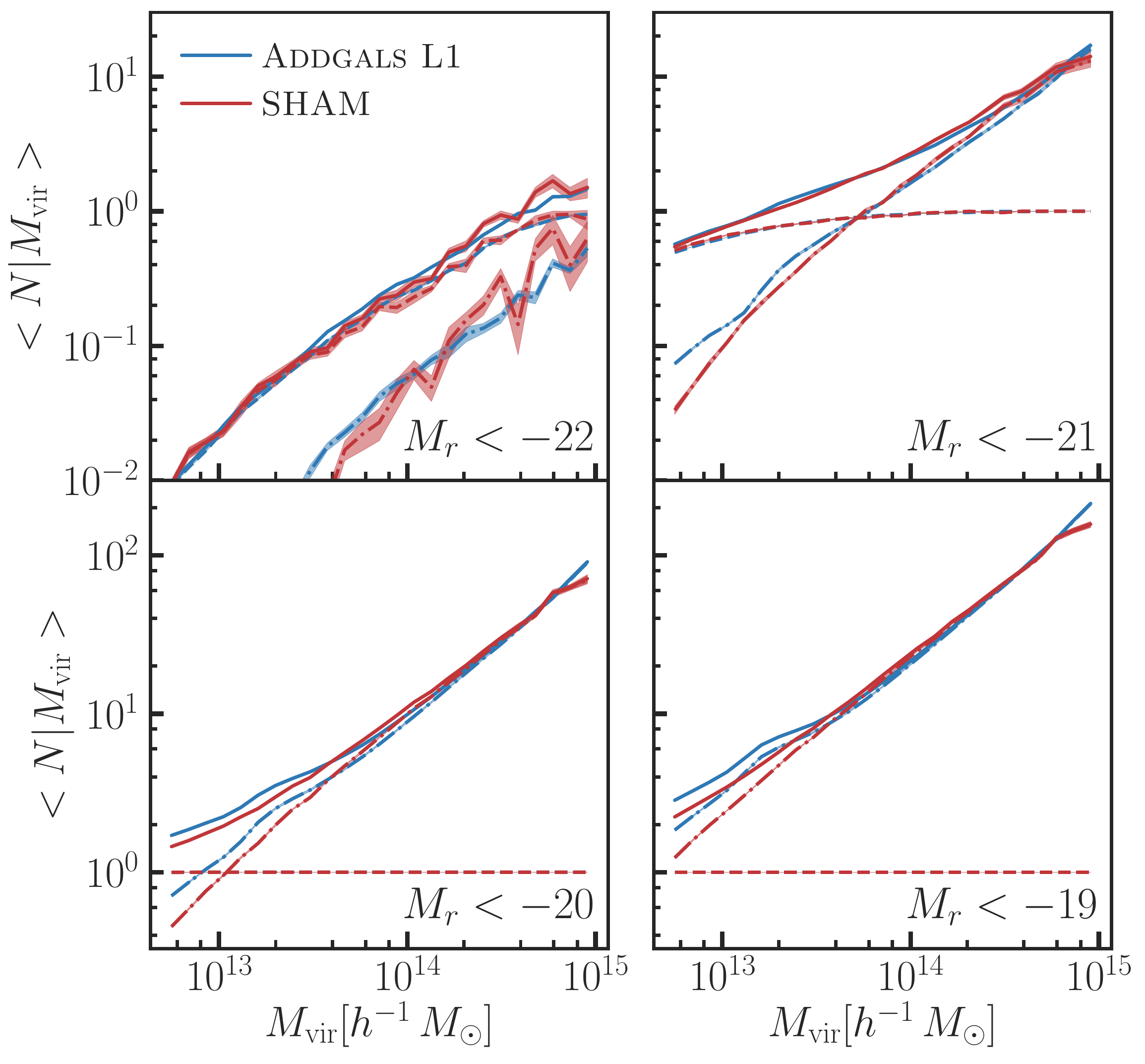}
\includegraphics[width=0.51\linewidth]{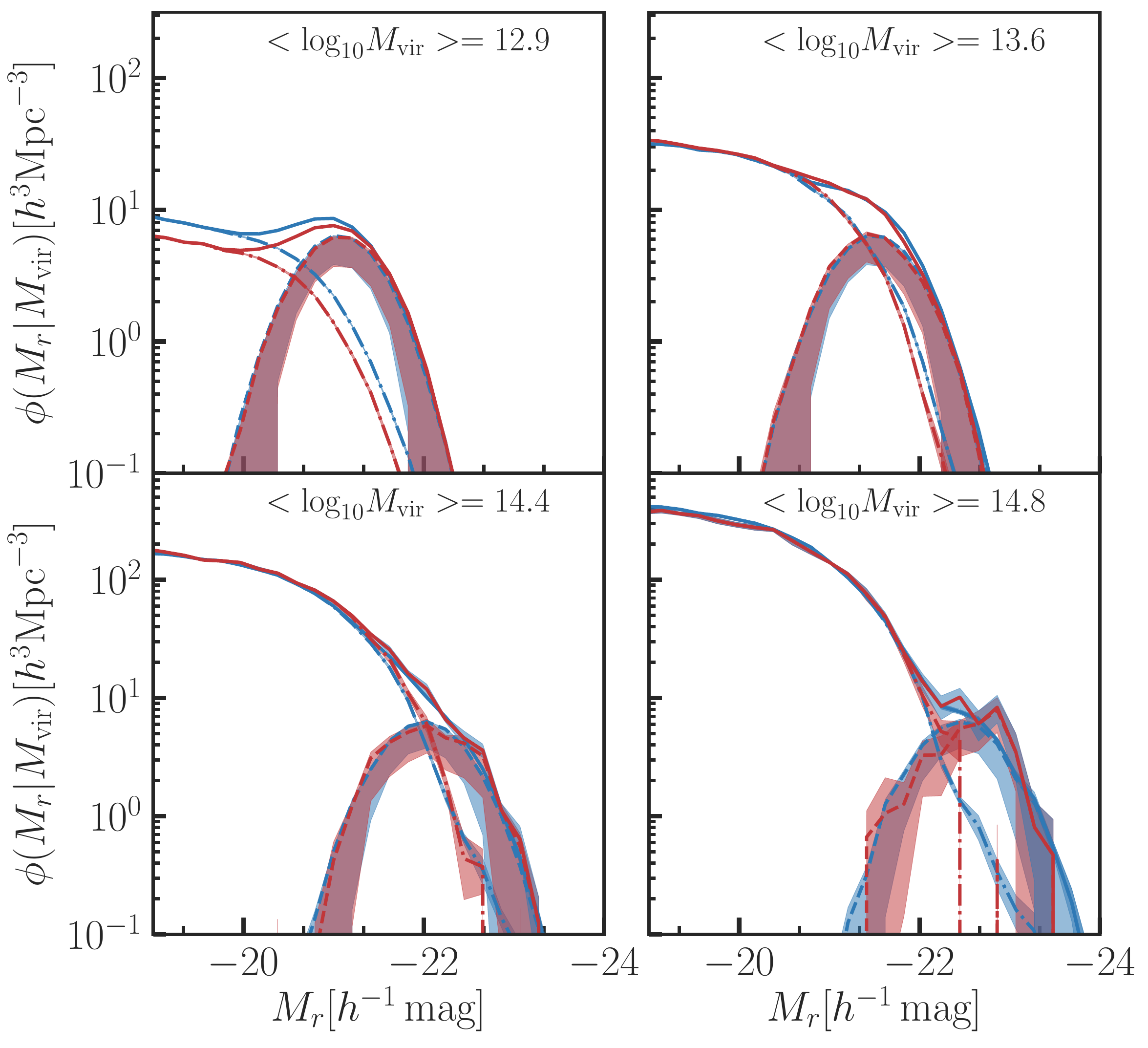}
\includegraphics[width=0.49\linewidth]{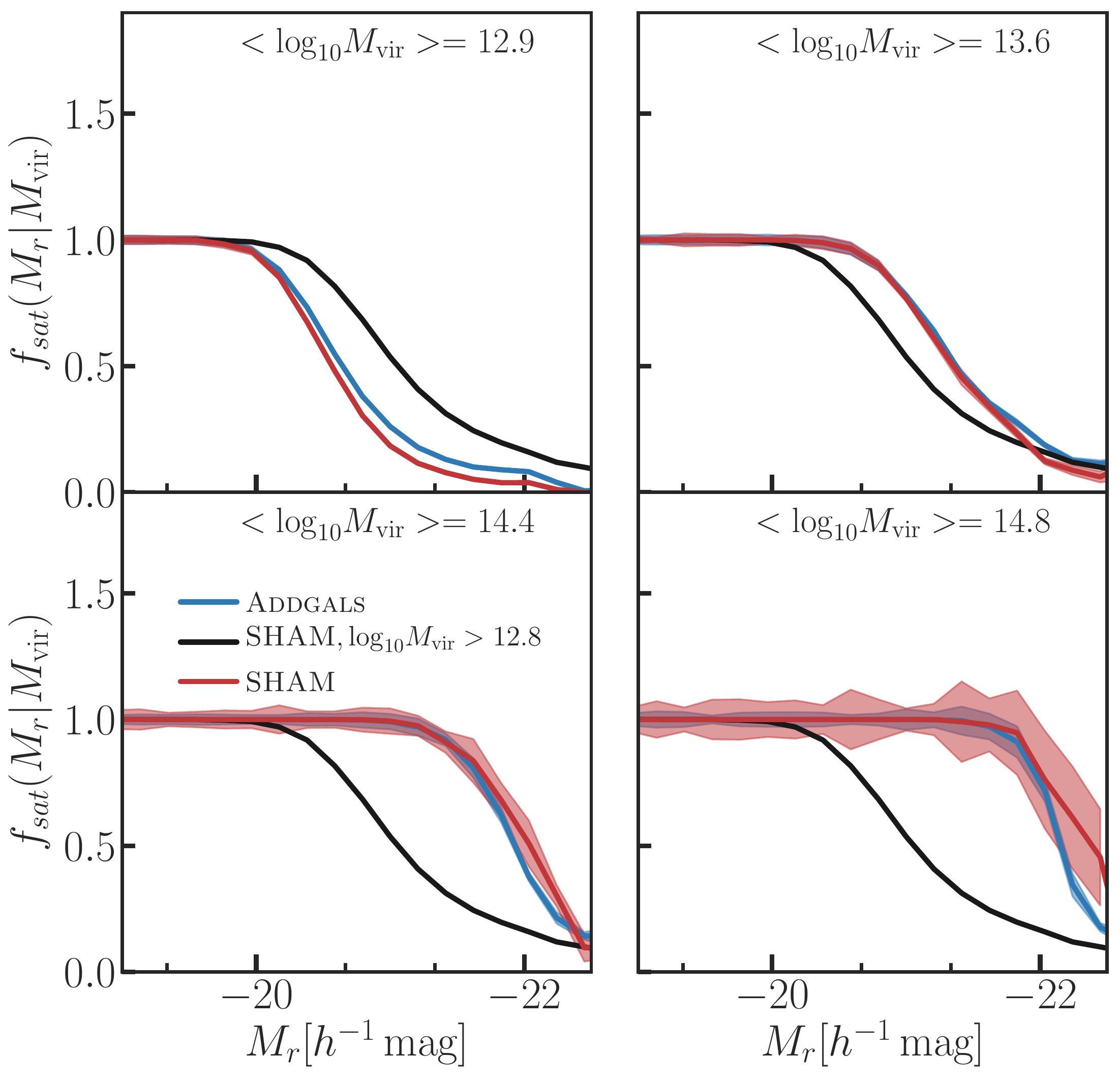}
\caption{
Comparison of halo occupation statistics between the \addgals\ L1 model (blue) and the SHAM model is it tuned to (red; based on the T1 simulation).
  \emph{Top left}: Central galaxy $r$-band absolute magnitude as a function of host halo mass. 
  \emph{Top right}: Halo occupation distribution for four magnitude-limited samples. Solid lines show total HODs, dashed lines show central HODs, and dash--dotted lines show satellite HODs. Error bars shown are the jackknife error bars for each catalog.
  \emph{Bottom left}: Conditional Luminosity Function for four bins in halo mass (the average  $\log(\hMsun)$ in the bin is labeled in each panel). Dashed lines are central luminosity functions, dash--dotted lines are satellite luminosity functions and solid lines are the sum of the two. Error bars indicate the jackknife error bars for each catalog.
  \emph{Bottom right}: Satellite fraction for galaxies in halos in four mass bins as a function of $r$-band absolute magnitude. For reference, he black line shows the satellite fraction in the \textsc{T1 SHAM} catalog for all halos with
  ${\rm log}_{10} M_{\rm vir}>12.8$.}
\label{fig:lumval}
\end{figure*}

Here we discuss the halo occupation statistics of the \addgals\ method compared to the SHAM model, applied to the T1 high-resolution training simulation.  The top left panel of \cref{fig:lumval} shows a comparison of the $p(M_{r,cen} | M_{halo})$ as measured in the $z=0$ snapshot of the \textsc{T1} SHAM catalog and the relation in an \addgals\ catalog run on the $z=0$ snapshot of the \textsc{L1} simulation. The agreement seen here is validation that the model in \cref{eq:bcg_ml} describes this relation well.
The small discrepancies seen at the bright end can be explained by differences in the assumed functional forms. The \addgals\ model for $M_{r, cen}(M_{\rm vir})$
  is a broken power law, which does not perfectly fit the relation measured in the
  SHAM model at the bright end. The luminosity function employed when creating the 
  SHAM catalog assumes a Schechter function plus a Gaussian component at the bright end,
  and the Gaussian component leads to a deviation from a power law for $M_r < -22$.
  The functional form for $M_{r, cen}(M_{\rm vir})$ used in \addgals\ was derived in 
  \citet{vale_ostriker:04} assuming a pure Schechter function, and is fit well by a pure
  power law for $M_r < -22$. This difference may give rise to slight discrepancies
  in the probability that the brightest galaxy in a cluster mass halo is the central galaxy 
  between the SHAM and \addgals\ models.

We compare the HODs and CLFs measured in the \textsc{SHAM} and \addgals\ catalogs. Since SHAM has been shown to provide a good match to the observed CLF \citep{Reddick12}, this comparison tests the assumption that the $p(\rd|M_r,z)$ relation is sufficient to recover a range of properties of the galaxy distribution and its relation to the underlying halos. In the top right panels of \cref{fig:lumval}, we compare the HOD as measured in the \textsc{Addgals L1} catalog and the \textsc{SHAM} catalog. \textsc{Addgals} largely agrees with the \textsc{SHAM} catalog, with some minor differences appearing around $10^{13} \hmsun$, where \addgals\ over-predicts abundances of galaxies with respect to \textsc{SHAM}. The reason for this is because at masses below the smoothing scale used to measure \rd( i.e. $1.8\times10^{13} h^{-1}M_{\odot}$) $P(\rd | M_r)$ becomes much broader, and thus is less able to disambiguate between halos of different masses. Due to the power-law halo mass function at low mass, galaxies that should have been placed in low-mass halos can then scatter into higher-mass halos, leading to the excess seen in the \addgals\ HOD measurements compared with the SHAM HODs.

The bottom left panel of \cref{fig:lumval} shows a comparison of the conditional luminosity function (CLF) of galaxies in bins of halo mass. Again, \addgals\ and the \textsc{SHAM} catalog are largely in agreement with each other, except for the lowest mass bin we consider, where a similar Eddington-like bias is at play. The bottom right panel of \cref{fig:lumval} shows a comparison of the fraction of galaxies that are satellites as a function of magnitude in the same mass bins as those used to measure the CLF. The satellite fraction for all galaxies in halos with $M_{\rm vir} > 5\times 10^{12} \hmsun$ is included in black for reference. At bright magnitudes, \addgals\ slightly over-predicts the satellite fraction in the lowest mass bin shown, and slightly under-predicts the bright-end satellite fraction for more massive halos, but these differences are small.

\end{appendices}

\bibliographystyle{apj}
\bibliography{addgals}

\end{document}